\documentclass[preprint,prd,showpacs,amsmath,amssymb]{revtex4}
\usepackage{bm}
\usepackage{graphicx}

\begin{document}
\title{Neutrino-Induced $\gamma$-Ray Emission from Supernovae}
\author{Yu Lu}
\email{ylu@physics.umn.edu}
\author{Yong-Zhong Qian}
\email{qian@physics.umn.edu}
\affiliation{School of Physics and Astronomy,
University of Minnesota, Minneapolis, MN 55455}
\date{\today}

\begin{abstract}
During a core-collapse supernova, absorption of $\bar\nu_e$ emitted 
from the proto-neutron star by protons in the hydrogen envelope produces
neutrons and positrons. Neutron capture on protons and positron annihilation
then produce $\gamma$ rays of 2.22 and 0.511~MeV, respectively. 
We calculate the fluxes of these $\gamma$ rays expected from a supernova 
with an $11\,M_\odot$ progenitor. 
The flux from neutron capture on protons exponentially 
decays on a timescale of 564~s, which is determined by neutron 
decay and capture on protons and $^3$He nuclei. The peak flux is
$2.38\times 10^{-7}\ {\rm cm}^{-2}\ {\rm s}^{-1}$ for a supernova at a distance
of 1~kpc. In contrast, the $\gamma$-ray flux from positron annihilation follows the 
time evolution of the $\bar\nu_e$ luminosity and lasts for $\sim 10$~s. 
The peak flux in this case is
$6.8\times10^{-5}\ {\rm cm}^{-2}\ {\rm s}^{-1}$ for a supernova at a distance
of 1~kpc. Detection of the above $\gamma$-ray fluxes is beyond the capability 
of current instruments, and perhaps even those planned for the near future. 
However,
if such fluxes can be detected, they not only constitute a new kind of signals that
occur during the gap of several hours between the neutrino signals and the 
optical display of a supernova,
but may also provide a useful probe of the conditions in the surface layers of
the supernova progenitor. 
\end{abstract}


\maketitle
                                                              
\section{Introduction}
\label{sec-intro}
On exhaustion of nuclear fuels, the core of a massive star 
($\gtrsim 8\,M_\odot$ with $M_\odot$ being the mass of the sun)
undergoes gravitational collapse, thereby initiating the supernova process.
Two classes of signals are expected from such an event: the neutrinos
emitted by the proto-neutron star formed from the collapsed core and
the photons radiated as the supernova shock emerges from the 
stellar surface. The neutrino signals start immediately after the shock is 
launched and last for $\sim 10$~s. However, it takes several hours for the
shock to emerge, and consequently, the associated photon radiation, 
in particular the optical display, is delayed from the neutrino burst by 
this shock propagation time. In this paper we consider a third class
of signals that occur before the shock emergence. These signals are
$\gamma$ rays induced by neutrino reactions in the stellar envelope.

We assume that the star undergoing core collapse still has its hydrogen
envelope. As the neutrinos from the proto-neutron star stream through
this envelope, the reaction
\begin{equation}
\bar\nu_e+p\to n+e^+
\label{eq-nup}
\end{equation}
produces a neutron and a positron. Subsequently, the capture of the 
neutron through
\begin{equation}
n+p\to {^2{\rm H}}+\gamma
\end{equation}
produces a $\gamma$ ray of 2.22~MeV, and the annihilation of the
positron can produce two $\gamma$ rays of 0.511 MeV each. 
The above mechanism
of $\gamma$-ray emission from supernovae has been considered earlier 
by Refs.~\cite{binc,rya}. However, these studies only estimated the expected 
$\gamma$-ray fluxes without giving an analysis of all the physical 
processes involved in the $\gamma$-ray production. For example, the
thermalization of neutrons and positrons in the stellar envelope was not 
discussed, and neither was the detailed time structure of the $\gamma$-ray 
emission. We note that $\gamma$ rays from neutron capture on protons
and positron annihilation were also discussed in the context of solar flares 
(see e.g., Ref.~\cite{hua}) and interstellar medium (see e.g., Ref.~\cite{buss}),
and there were extensive studies of the physical processes involved in
the $\gamma$-ray emission from positron annihilation 
(see e.g., Ref.~\cite{murphy}).

We here present detailed analyses of the major physical processes that
lead to $\gamma$-ray production following the reaction 
$\bar\nu_e+p\to n+e^+$ in the hydrogen envelope of a massive star.
In particular, we show that the $\gamma$-ray emission due to neutron
capture on protons lasts for $\sim 10^3$~s while that due to positron annihilation
follows the time evolution of the $\bar\nu_e$ luminosity and lasts for $\sim 10$~s. 
For concreteness,
we adopt a specific model of neutrino emission and a specific stellar model 
for the conditions in the hydrogen envelope. Our analyses can be easily 
generalized to other neutrino emission and stellar models. 

We assume that the gravitational binding energy 
$E_B$ of the final neutron star is emitted equally in 
$\nu_e$, $\bar\nu_e$, $\nu_\mu$, $\bar\nu_\mu$, $\nu_\tau$, and 
$\bar\nu_\tau$ and that the $\bar\nu_e$ luminosity $L_{\bar\nu_e}(t)$
decays exponentially on a timescale $\tau$. Thus,
\begin{equation}
L_{\bar\nu_e}(t)=\frac{E_B}{6\tau}\exp(-t/\tau).
\end{equation}
We take $E_B=3\times 10^{53}$~erg and $\tau=3$~s. The normalized
$\bar\nu_e$ energy spectrum is taken to be
\begin{equation}
f_{\bar\nu_e}(E_{\bar\nu_e})=\frac{1}{T_{\bar\nu_e}^3F_2(\eta_{\bar\nu_e})}
\frac{E_{\bar\nu_e}^2}{\exp[(E_{\bar\nu_e}/T_{\bar\nu_e})-\eta_{\bar\nu_e}]+1},
\label{eq-nuspec}
\end{equation}
where $T_{\bar\nu_e}=3.76$~MeV, $\eta_{\bar\nu_e}=3$, and
\begin{equation}
F_n(\eta)\equiv\int_0^\infty\frac{x^n}{\exp(x-\eta)+1}dx.
\end{equation}
The corresponding average $\bar\nu_e$ energy is
$\langle E_{\bar\nu_e}\rangle=T_{\bar\nu_e}
F_3(\eta_{\bar\nu_e})/F_2(\eta_{\bar\nu_e})=15$~MeV. The cross section
for the reaction $\bar\nu_e+p\to n+e^+$ (see Appendix~\ref{sec-xs})
averaged over the above spectrum is
$\langle\sigma_{\bar\nu_ep}\rangle=1.87\times 10^{-41}\ {\rm cm}^2$.

For the conditions of the hydrogen envelope, we use the model of an 
$11\,M_\odot$ star in Ref.~\cite{rito}. 
The region of interest for $\gamma$-ray emission is limited by the interaction
of $\gamma$ rays with matter. For $\gamma$ rays of $\sim 1$~MeV produced
in the hydrogen envelope, the dominant interaction is Compton scattering on 
electrons. The relevant cross sections (see Appendix~\ref{sec-xs}) are 
$\sigma_{\gamma(np)e}=1.38\times 10^{-25}$~cm$^2$ and 
$\sigma_{\gamma(e^\pm)e}=2.87\times 10^{-25}\ {\rm cm}^2$ for $\gamma$ 
rays of 2.22 and 0.511~MeV, respectively. The corresponding mean free path is
\begin{equation}
l_{\gamma e}=\frac{1}{n_e\sigma_{\gamma e}}=1.66\times10^9
\left(\frac{10^{-8}\ \mathrm{g\ cm}^{-3}}{\rho Y_e}\right)
\left(\frac{10^{-25}\ \mathrm{cm}^2}{\sigma_{\gamma e}}\right)\ \mathrm{cm},
\end{equation}
where $n_e=\rho Y_eN_A$ is the electron number density, $\rho$ is the matter
density, $Y_e$ is the number of electrons per nucleon, and $N_A$ is Avogadro's
number. The surface zone of our adopted stellar model has 
$\rho=1.59\times 10^{-8}$~g~cm$^{-3}$ and $Y_e=0.85$, for which
$l_{\gamma(np)e}=8.9\times 10^8$~cm and 
$l_{\gamma(e^\pm)e}=4.28\times 10^8$~cm for $\gamma$ rays 
of 2.22 and 0.511~MeV, respectively. The radius of this zone is 
$R=2.36\times 10^{13}$~cm. For considering the emission of 2.22 and 
0.511~MeV $\gamma$ rays (we do not treat the emission of scattered
$\gamma$ rays at other energies), it is sufficient to focus on the outermost region
with a radial thickness $d$ satisfying $l_{\gamma e}\ll d\ll R$.
The stellar conditions stay constant in this region. In addition,
the rate for production of neutrons and positrons 
per nucleon by the reaction $\bar\nu_e+p\to n+e^+$ is the
same throughout this region and is
\begin{equation}
\lambda_{\bar\nu_ep}(t)=Y_p\left(\frac{\langle\sigma_{\bar\nu_ep}\rangle}
{4\pi R^2}\right)\frac{L_{\bar\nu_e}(t)}{\langle E_{\bar\nu_e}\rangle}
=\lambda_{\bar\nu_ep}(0)\exp(-t/\tau),
\label{eq-rnup}
\end{equation}
where $Y_p=0.7$ is the number of protons per nucleon in the region, and
$\lambda_{\bar\nu_ep}(0)=1.30\times 10^{-12}$~s$^{-1}$ for the adopted 
parameters.

We study the emission of 2.22~MeV $\gamma$ rays from neutron capture
on protons in Sec.~\ref{sec-ncap}  and that of 0.511~MeV $\gamma$ rays 
from positron
annihilation in Sec.~\ref{sec-pos}. We discuss our results and give
conclusions in Sec.~\ref{sec-dis}.

\section{$\gamma$-Ray Emission from Neutron Capture on Protons}
\label{sec-ncap}
Before discussing the physical processes leading to $\gamma$-ray emission 
from neutron capture on protons, we give a simple estimate of the time evolution 
of the
corresponding flux. This evolution is closely related to that of the neutron number
per nucleon $Y_n(t)$ in the stellar surface region from which $\gamma$ rays 
can escape efficiently. The increase in $Y_n(t)$ is due to $\bar\nu_e$ absorption
by protons with the rate $\lambda_{\bar\nu_ep}(t)$ given in Eq.~(\ref{eq-rnup}). 
The decrease in $Y_n(t)$ is caused by neutron decay and capture onto protons 
and $^3$He. The neutron lifetime is $\tau_n=887$~s. For capture of low-energy 
neutrons, the cross section is inversely proportional to the neutron velocity $v_n$.
Consequently, the product of the cross section and $v_n$ is independent of the
neutron velocity distribution. We use $\langle v_n\sigma_{np}\rangle=
7.32\times 10^{-20}$~cm$^3$~s$^{-1}$ for capture onto protons and
$\langle v_n\sigma_{n3}\rangle=1.17\times 10^{-15}$~cm$^3$~s$^{-1}$ for 
capture onto $^3$He. The corresponding capture timescales are
$\tau_{np}=(\rho Y_pN_A\langle v_n\sigma_{np}\rangle)^{-1}=2.04\times 10^3$~s 
and
$\tau_{n3}=(\rho Y_3N_A\langle v_n\sigma_{n3}\rangle)^{-1}=6.38\times 10^3$~s,
where we have taken the number of $^3$He per nucleon to be 
$Y_3=1.4\times 10^{-5}$, the same as estimated for the solar photosphere
\cite{hua}. Note that the dominant channel of neutron capture onto $^3$He 
is $n+{^3{\rm He}}\to p+{^3{\rm H}}$, which does not produce any $\gamma$ ray.
However, the enormous cross section of this channel enables it to play a 
significant role in determining the evolution of $Y_n(t)$ in spite of the small 
abundance of $^3$He.

A simple estimate of $Y_n(t)$ can be obtained from
\begin{equation}
\frac{dY_n}{dt}=\lambda_{\bar\nu_ep}(t)-\frac{Y_n(t)}{\tau_{\rm eff}},
\end{equation}
where $\tau_{\rm eff}^{-1}=\tau_n^{-1}+\tau_{np}^{-1}+\tau_{n3}^{-1}$ and
$\tau_{\rm eff}=564$~s. The solution to the above equation is
\begin{equation}
Y_n(t)=\lambda_{\bar\nu_ep}(0)\tau
\left(\frac{\tau_{\rm eff}}{\tau_{\rm eff}-\tau}\right)
[\exp(-t/\tau_{\rm eff})-\exp(-t/\tau)].
\end{equation}
As $\tau_{\rm eff}\gg\tau$, $Y_n(t)$ rises to its peak value
\begin{equation}
Y_n^{\rm pk}\approx\lambda_{\bar\nu_ep}(0)\tau=3.90\times 10^{-12}
\label{eq-ynpk}
\end{equation}
on a timescale of $\sim 10$~s and then exponentially decays on the timescale
$\tau_{\rm eff}$.

Due to Compton scattering, only those $\gamma$-rays of 2.22 MeV produced
in the outermost stellar layer with a thickness of
$\sim l_{\gamma(np)e}=8.9\times 10^8$~cm will escape efficiently.
The corresponding flux at a radius $r>R$ can be estimated as
\begin{equation}
\Phi_{\gamma(np)}(r,t_r)\sim\frac{\rho Y_n(t)N_AR^2l_{\gamma(np)e}}{\tau_{np}r^2}
\sim 9.53\times 10^{-7}\left(\frac{1\ {\rm kpc}}{r}\right)^2\exp(-t/\tau_{\rm eff})\ 
{\rm cm}^{-2}\ {\rm s}^{-1},
\label{eq-fnpa}
\end{equation}
where $t_r$ is the time at which the $\gamma$ rays emitted at time $t$ 
arrive at radius $r$
and we have used $Y_n(t)\sim Y_n^{\rm pk}\exp(-t/\tau_{\rm eff})$ in the second
approximation. To show the dependence on the model of neutrino emission
and stellar conditions, we rewrite the above equation as
\begin{equation}
\Phi_{\gamma(np)}(r,t_r)\sim\frac{E_B}{24\pi r^2\langle E_{\bar\nu_e}\rangle\tau_{np}}
\left(\frac{Y_p\langle\sigma_{\bar\nu_ep}\rangle}{Y_e\sigma_{\gamma(np)e}}\right)
\exp(-t/\tau_{\rm eff}).
\end{equation}
It can be seen that the exact form of $L_{\bar\nu_e}(t)$ is unimportant so long as
the timescale of neutrino emission is $\ll\tau_{\rm eff}$. In addition, the density of
the stellar surface region controls the peak magnitude and the decay timescale
of the flux via $\tau_{np}$ and $\tau_{\rm eff}$, respectively. Integrating the flux
over time, we estimate the total fluence of 2.22 MeV $\gamma$ rays at radius 
$r$ as
\begin{equation}
{\cal{F}}_{\gamma(np)}\sim\frac{E_B}{24\pi r^2\langle E_{\bar\nu_e}\rangle}
\left(\frac{Y_p\langle\sigma_{\bar\nu_ep}\rangle}{Y_e\sigma_{\gamma(np)e}}\right)
\left(\frac{\tau_{\rm eff}}{\tau_{np}}\right)
\sim 5.37\times 10^{-4}\left(\frac{1\ {\rm kpc}}{r}\right)^2\ {\rm cm}^{-2}.
\end{equation}
Note that ${\cal{F}}_{\gamma(np)}$ in general still depends on the density of the 
stellar surface region due to the competition between neutron decay and capture.
This dependence ceases only when neutron decay can be ignored
(i.e., $\tau_n\gg\tau_{np}$). 

\subsection{Thermalization and Diffusion of Neutrons}
As discussed above, the timescales for neutron capture onto protons and $^3$He
are $\tau_{np}=2.04\times 10^3$~s and $\tau_{n3}=6.38\times 10^3$~s for the 
adopted density and composition of the stellar surface region. We now show that
these timescales and the neutron lifetime are so long that neutrons are thermalized 
due to scattering by protons before being captured or decay. 
Following the absorption of $\bar\nu_e$ by protons, 
positrons are emitted approximately isotropically with energies of 
$\approx E_{\bar\nu_e}$. Using the $\bar\nu_e$ energy
spectrum in Eq.~(\ref{eq-nuspec}) and the cross section in Eq.~(\ref{eq-nux}), 
we obtain the average recoil energy of the neutrons produced along with the 
positrons as $\langle E_n^{\rm rec}\rangle\sim 
[F_6(\eta_{\bar\nu_e})/F_4(\eta_{\bar\nu_e})]T_{\bar\nu_e}^2/M_n\sim 543$~keV, 
where $M_n$ is the neutron rest mass. The region of interest has a temperature
$T=1.53\times 10^4$~K corresponding to a thermal energy 
$E_{\rm th}=(3/2)kT=1.98$~eV~$\ll\langle E_n^{\rm rec}\rangle$ with $k$ being
Boltzmann's constant. Consequently, neutrons lose energy through scattering 
by thermal protons until neutrons are thermalized (scattering by other particles 
can be ignored). The average decrease in the natural logarithm of neutron energy 
per scattering is unity \cite{segre-nsc}:
$\langle\ln(E_{n,j+1}/E_{n,j})\rangle=-1$, where $E_{n,j}$ is the neutron energy
after $j$ scatterings. So we approximately have
\begin{equation}
E_{n,j}\sim\langle E_n^{\rm rec}\rangle\exp(-j).
\end{equation}
The mean free path between scattering is
\begin{equation}
l_{\rm sc}=\frac{1}{n_p\sigma_{\rm sc}}=1.66\times 10^7
\left(\frac{10^{-8}\ {\rm g\ cm}^{-3}}{\rho Y_p}\right)
\left(\frac{10\ {\rm b}}{\sigma_{\rm sc}}\right)\ {\rm cm},
\end{equation}
where $n_p=\rho Y_pN_A$ is the proton number density and
$\sigma_{\rm sc}$ is the scattering cross section. For the relevant
neutron energies, $\sigma_{\rm sc}\sim 10$~b \cite{segre-nx} corresponds
to $l_{\rm sc}\sim 1.49\times 10^7$~cm for the adopted stellar conditions.
The timescale for thermalization can then be estimated as
\begin{equation}
\tau_{\rm therm}\sim\sum_{j=0}^{j_{\rm max}-1}
\frac{l_{\rm sc}}{\sqrt{2E_{n,j}/M_n}}\sim 15\ {\rm s}
\ll\tau_n,\ \tau_{np},\ \tau_{n3},
\end{equation}
where $j_{\rm max}\sim\ln(\langle E_n^{\rm rec}\rangle/E_{\rm th})\sim 13$.

Once thermalized, neutrons diffuse until they decay or are captured or escape
from the stellar surface. The mean speed of thermal neutrons is
$\bar v_n=\sqrt{8kT/(\pi M_n)}=1.80\times 10^6$~cm~s$^{-1}$. If they escape
from the star under consideration ($11\,M_\odot$ with a radius 
$R=2.36\times 10^{13}$~cm), they will not fall back onto the star as they
will decay while they are still moving away from the star.
Therefore, those neutrons that can escape will not contribute to the production 
of $\gamma$ rays. In order to escape, neutrons must diffuse to the stellar 
surface on timescales shorter than $\tau_{\rm eff}$. The thickness $\delta$ of 
the layer from which neutrons can diffuse to escape can be estimated from 
$\delta^2\sim (\bar v_n\tau_{\rm eff}/l_{\rm sc,th})l_{\rm sc,th}^2$, which gives
$\delta\sim\sqrt{\bar v_n\tau_{\rm eff}l_{\rm sc,th}}=8.6\times 10^7$~cm with
$l_{\rm sc,th}=7.28\times 10^6$~cm corresponding to $\sigma_{\rm sc,th}=20.5$~b
for thermal neutrons. As $\delta\sim 0.1l_{\gamma(np)e}$, the reduction of
potential $\gamma$-ray production due to the escape of neutrons is insignificant.
For the outmost layer of thickness $l_{\gamma(np)e}$ from which 
$\gamma$ rays can escape efficiently, essentially all the neutrons in this layer 
decay or are captured by protons or $^3$He during diffusion. Indeed, the timescale 
for neutrons to diffuse out of this layer is
\begin{equation}
\tau_{\rm diff}=\frac{l_{\gamma(np)e}^2}{l_{\rm sc,th}\bar v_n}=6.04\times 10^4\ {\rm s}
\gg\tau_{\rm eff}.
\label{eq-tdiff}
\end{equation}
For the adopted stellar conditions, $\tau_n=\tau_{np}/2.3$ and 
$\tau_{n3}=3.1\tau_{np}$. So a fraction $\tau_{\rm eff}/\tau_{np}=
[1+(\tau_{np}/\tau_n)+(\tau_{np}/\tau_{n3})]^{-1}=27.6\%$ of the neutrons
are captured by protons to produce 2.22~MeV $\gamma$ rays. The reduction of
potential $\gamma$-ray production is mainly due to neutron decay and to a smaller
extent due to neutron capture by $^3$He. 
The rates and mean free paths for the important processes 
involved in the $\gamma$-ray emission from
neutron capture on protons in the star under consideration
are summarized in Table~\ref{tab-ncap}.

\begin{table}
\caption{Rate, mean free paths, and timescales for the important processes 
involved in the $\gamma$-ray emission from neutron capture on protons. 
The adopted conditions in the stellar surface region are characterized by
$R=2.36\times 10^{13}$~cm,
$\rho=1.59\times 10^{-8}\ {\rm g}\ {\rm cm}^{-3}$, $Y_e=0.85$, $Y_p=0.7$,  
and $Y_3=1.4\times 10^{-5}$.
\label{tab-ncap}}
\begin{ruledtabular}
\begin{tabular}{cccccc}
$\lambda_{\bar\nu_ep}(0)$&$l_{\gamma(np)e}$&$l_{\rm sc,th}$&$\tau_{np}$&
$\tau_{n3}$&$\tau_{\rm eff}$\\\hline
$1.30\times 10^{-12}\ {\rm s}^{-1}$&$8.9\times 10^8$~cm&$7.28\times 10^6$~cm&
$2.04\times 10^3$~s&$6.38\times 10^3$~s&564 s\\
\end{tabular}
\end{ruledtabular}
\end{table}

\subsection{Evolution of Neutron Density and Emergent $\gamma$-Ray Flux}
\begin{figure}
\includegraphics[scale=0.3]{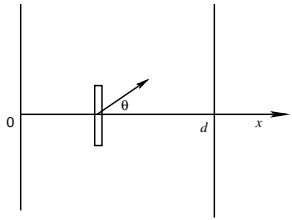}%
\caption{Sketch of the stellar surface region of interest to $\gamma$-ray
emission. As $l_{\gamma e}\ll d\ll R$, this region can be treated as a slab
of thickness $d$ perpendicular to the $x$-axis with the surface at $x=d$.
The calculation of the local $\gamma$-ray flux in the positive $x$-direction
involves integration over the forward solid angle defined by 
$0\leq\theta\leq\pi/2$.
\label{fig-slab}}
\end{figure}

Based on the preceding discussion, we consider the following quantitative model
for calculating the flux of 2.22~MeV $\gamma$ rays from the star
under consideration.  We focus on the outermost layer of thickness 
$d=10^{10}\ {\rm cm}\gg l_{\gamma(np)e}$ to find the emergent flux 
$\Phi_{\gamma(np)}(R,t)$ at the stellar surface. 
As $d\ll R$, we can treat the layer of interest as a slab perpendicular to the 
$x$-axis with the stellar surface at $x=d$ (see Fig.~\ref{fig-slab}). Knowing
the neutron number per nucleon $Y_n(x,t)$ in this layer, we can calculate the
local flux of 2.22~MeV $\gamma$ rays in the positive $x$-direction as
\begin{equation}
\frac{Y_n(x,t)\rho N_A}{4\pi\tau_{np}}
\int_0^{\pi/2}\cos\theta\sin\theta d\theta\int_0^{2\pi}d\phi=
\frac{Y_n(x,t)\rho N_A}{4\tau_{np}},
\label{eq-floc}
\end{equation}
where $\theta$ and $\phi$ are respectively, the polar and azimuthal angles 
with respect to the $x$-axis. Consequently, the emergent flux at the stellar
surface is
\begin{equation}
\Phi_{\gamma(np)}(R,t)=\frac{\rho N_A}{4\tau_{np}}
\int_0^d Y_n(x,t)\exp\left(-\frac{d-x}{l_{\gamma(np)e}}\right)dx,
\label{eq-fem}
\end{equation}
which is related to the flux at $r>R$ as 
$\Phi_{\gamma(np)}(r,t_r)=(R/r)^2\Phi_{\gamma(np)}(R,t)$. 

\begin{figure}
\includegraphics[scale=0.25,angle=270]{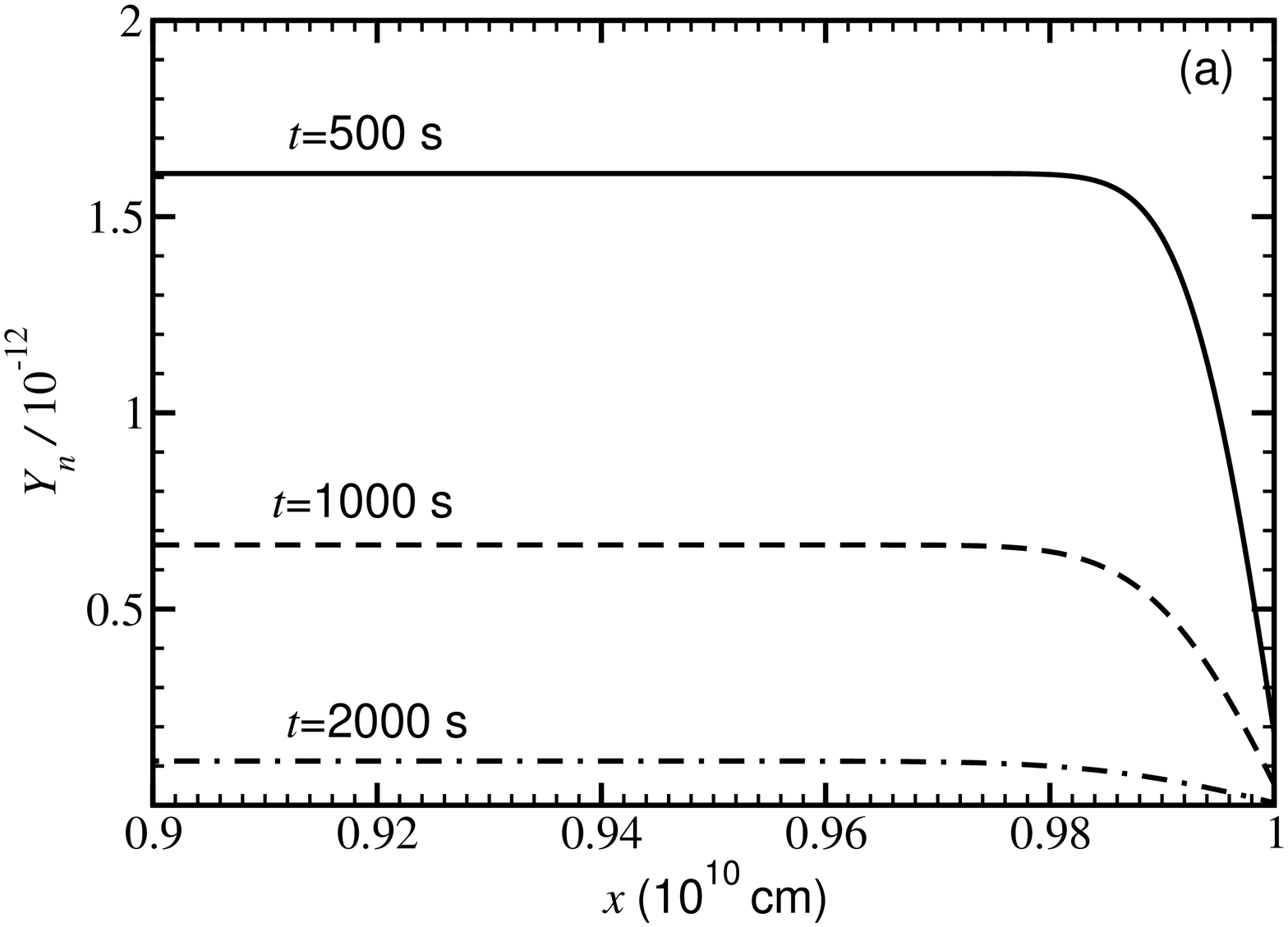}%
\includegraphics[scale=0.25,angle=270]{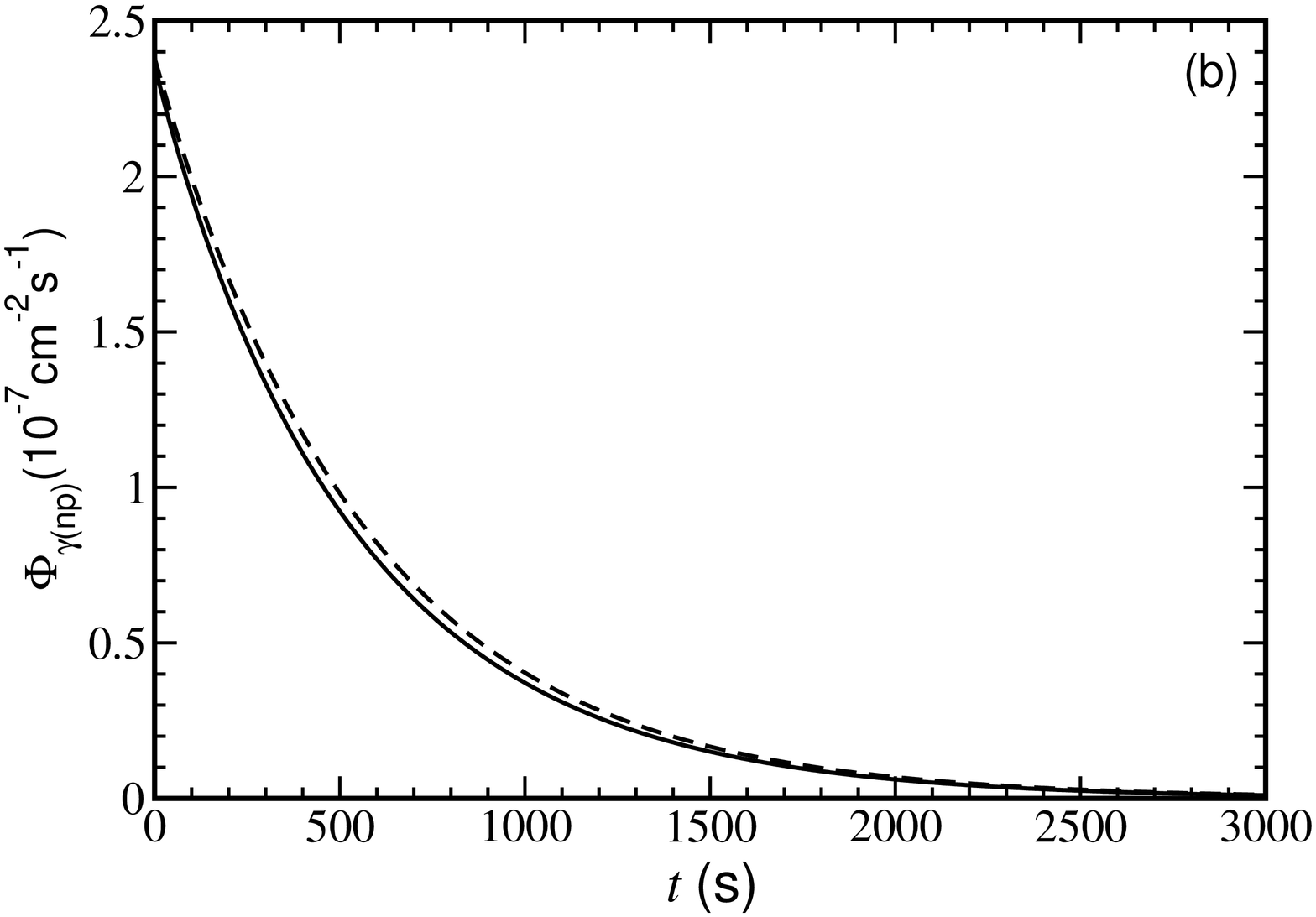}%
\caption{(a) Neutron number per nucleon $Y_n(x,t)$ as functions of $x$
in the stellar surface region for $t=500$, 1000, and 2000~s. Note that
the decline of $Y_n(x,t)$ with $x$ due to diffusion only occurs in a
narrow region near the surface. (b) Time evolution for the expected flux 
of 2.22~MeV $\gamma$ rays from neutron capture on protons in a 
supernova at a distance of 1~kpc as calculated by solving the diffusion 
equation (solid curve) and estimated by neglecting diffusion (dashed curve).
\label{fig-fnp}}
\end{figure}

The evolution of
$Y_n(x,t)$ is governed by diffusion:
\begin{equation}
\frac{\partial Y_n}{\partial t}=D\frac{\partial^2 Y_n}{\partial x^2}-
\frac{Y_n(x,t)}{\tau_{\rm eff}},
\label{eq-ydiff}
\end{equation}
where $D=l_{\rm sc,th}\bar v_n/3$ is the diffusion coefficient of thermal neutrons.
As $\tau_{\rm eff}$ is much longer than the timescales for neutrino emission 
and thermalization of neutrons, we assume $Y_n(x,0)=Y_n^{\rm pk}$ 
[see Eq.~(\ref{eq-ynpk})]. Further, on the timescale relevant for $\gamma$-ray
production, diffusion has little effect on the spatial 
distribution of neutrons at $x=0$ [see Eq.~(\ref{eq-tdiff})]. So we take 
$(\partial Y_n/\partial x)_{x=0}=0$ as the inner boundary condition.
For the outer boundary at $x=d$, we assume that the drift neutron flux given by 
diffusion is the same as the escaping neutron flux:
$-D(\partial Y_n/\partial x)_{x=d}=\bar v_nY_n(d,t)/4$, which is equivalent to
$l_{\rm sc,th}(\partial Y_n/\partial x)_{x=d}=-(3/4)Y_n(d,t)$. 
With the initial and boundary conditions discussed above, the solution to
Eq.~(\ref{eq-ydiff}) is
\begin{equation}
Y_n(x,t)=2Y_n^{\rm pk}\sum_{j=0}^{\infty}
\frac{\sin(\kappa_jd)\cos(\kappa_jx)}{\kappa_jd+\sin(\kappa_jd)\cos(\kappa_jd)}
\exp[-(\kappa_j^2D+\tau_{\rm eff}^{-1})t],
\label{eq-yn}
\end{equation}
where $\kappa_j$ satisfies $\kappa_jl_{\rm sc,th}\tan(\kappa_jd)=3/4$ as 
required by the outer boundary condition. The evolution of $Y_n(x,t)$ given by 
Eq.~(\ref{eq-yn}) is shown in Fig.~\ref{fig-fnp}a.

Substituting the above expression of $Y_n(x,t)$ in Eq.~(\ref{eq-fem}), we
obtain
\begin{equation}
\Phi_{\gamma(np)}(R,t)=
\frac{Y_n^{\rm pk}\rho N_Al_{\gamma(np)e}}{2\tau_{np}}
\sum_{j=0}^{\infty}A_j\exp[-(\kappa_j^2D+\tau_{\rm eff}^{-1})t]
\label{eq-fnpx}
\end{equation}
where
\begin{equation}
A_j=\frac{\sin(\kappa_jd)[\cos(\kappa_jd)+
(\kappa_jl_{\gamma(np)e})\sin(\kappa_jd)-\exp(-d/l_{\gamma(np)e})]}
{[\kappa_jd+\sin(\kappa_jd)\cos(\kappa_jd)][1+(\kappa_jl_{\gamma(np)e})^2]}.
\label{eq-aj}
\end{equation}
As $l_{\rm sc,th}\ll l_{\gamma(np)e}\ll d$, it can be shown that
\begin{equation}
A_j\approx\frac{l_{\gamma(np)e}/d}{[1+(4\kappa_jl_{\rm sc,th}/3)^2]
[1+(\kappa_jl_{\gamma(np)e})^2]}.
\end{equation}
The outer boundary condition gives $j\pi/d<\kappa_j<[(j+(1/2)]\pi/d$.
For $j\ll 3d/(4\pi l_{\rm sc,th})=328$, $\kappa_j\approx [j+(1/2)]\pi/d$.
This approximation can be used for all $j$ as there is little difference 
between $j$ and $j+(1/2)$ for sufficiently large $j$. Noting that
$4\kappa_jl_{\rm sc,th}/3\approx[j+(1/2)]/328$,
$\kappa_jl_{\gamma(np)e}\approx[j+(1/2)]/3.58$, and
$\kappa_j^2D\tau_{\rm eff}\approx\{[j+(1/2)]/64.1\}^2$, we have
\begin{eqnarray}
\Phi_{\gamma(np)}(r,t_r)&=&\left(\frac{R}{r}\right)^2\Phi_{\gamma(np)}(R,t)
\nonumber\\
&\approx&\frac{Y_n^{\rm pk}\rho N_AR^2l_{\gamma(np)e}}{2\tau_{np}r^2}
\exp(-t/\tau_{\rm eff})\sum_{j=0}^{\infty}
\frac{l_{\gamma(np)e}/d}{1+\{[j+(1/2)]\pi l_{\gamma(np)e}/d\}^2}
\label{eq-fnpsum}\\
&\approx&2.38\times 10^{-7}\left(\frac{1\ {\rm kpc}}{r}\right)^2
\exp(-t/\tau_{\rm eff})\ {\rm cm}^{-2}\ {\rm s}^{-1}.
\label{eq-fnpb}
\end{eqnarray}
The result in Eq.~(\ref{eq-fnpb}) is obtained by approximating the sum in 
Eq.~(\ref{eq-fnpsum}) by the integral $\int_0^\infty(1+y^2)^{-1}dy/\pi=1/2$.
This result is smaller than the estimate in Eq.~(\ref{eq-fnpa})
by a factor of 4, which comes from the integration over the solid
angle to obtain the local flux at a point in the region of $\gamma$-ray
production [see Eq.~(\ref{eq-floc})]. We numerically evaluate
the flux of 2.22~MeV $\gamma$ rays
at $r=1$~kpc as a function of time $t$
from Eqs.~(\ref{eq-fnpx}) and (\ref{eq-aj}) and show the result in 
Fig.~\ref{fig-fnp}b along with the approximation in Eq.~(\ref{eq-fnpb}).
The numerical result is essentially the same as the approximation.

\section{$\gamma$-Ray Emission from Positron Annihilation}
\label{sec-pos}
The processes leading to the emission of 0.511~MeV $\gamma$ rays from 
annihilation of positrons subsequent to their production by $\bar\nu_e$ 
absorption on protons are much more complicated than those resulting in
the emission of 2.22~MeV $\gamma$ rays from neutron capture on 
protons. This is 
because a positron can lose energy through many processes such as
electronic excitations of ionized and atomic matter, bremsstrahlung, 
Compton scattering, and synchrotron radiation in the presence of a
magnetic field. In addition, a positron can directly annihilate with or form
a positronium (Ps) with a free or bound electron:
\begin{eqnarray}
e^++e^-&\to&\gamma+\gamma,\label{eq-daf}\\
e^++{\rm H}&\to&{\rm H}^++\gamma+\gamma,\\
e^++{\rm He}&\to&{\rm He}^++\gamma+\gamma,\\
e^++e^-&\to&{\rm Ps}+\gamma,\\
e^++{\rm H}&\to&{\rm Ps}+{\rm H}^+,\label{eq-psh}\\
e^++{\rm He}&\to&{\rm Ps}+{\rm He}^+.\label{eq-pshe}
\end{eqnarray}
In the case of Ps formation (with free and bound electrons through 
radiative combination and charge exchange, respectively), 
the singlet Ps ($^1$Ps with total spin 0) decays by emitting two 
photons of 0.511~MeV each while the triplet Ps ($^3$Ps with total spin 1) 
decays by emitting three photons with a continuous energy spectrum. 
Here we focus on the production of 0.511~MeV
$\gamma$ rays only. We will loosely refer to both direct annihilation of
positrons and Ps decay as ``annihilation'' of positrons.

The energy loss and the annihilation processes of the positrons depend
on the ionization state of the medium. We consider
that the stellar surface region is in local thermodynamic equilibrium at
$T=1.53\times 10^4$~K and $\rho=1.59\times 10^{-8}$~g~cm$^{-3}$.
We assume that the forward and reverse processes of the following 
chemical reactions are in equilibrium:
\begin{eqnarray}
e^-+{\rm H}^+&\rightleftharpoons&{\rm H}+\gamma,
\label{eq-eh}\\
e^-+{\rm He}^+&\rightleftharpoons&{\rm He}+\gamma.
\label{eq-ehe}
\end{eqnarray}
As we will see shortly, nearly all of the He atoms remain neutral and
therefore, we can ignore the presence of He$^{++}$. To avoid confusion,
we denote the numbers of free $e^-$, H$^+$ (free protons), 
He$^+$, and neutral H and 
He atoms per nucleon as $Y_{e^-}$,  $Y_{{\rm H}^+}$, $Y_{{\rm He}^+}$,
$Y_{\rm H}$, and $Y_{\rm He}$, respectively. These quantities satisfy
\begin{eqnarray}
Y_{{\rm H}^+}+Y_{\rm H}&=&Y_p,\label{eq-yh}\\
Y_{{\rm He}^+}+Y_{\rm He}&=&Y_\alpha,\label{eq-yhe}\\
Y_{{\rm H}^+}+Y_{{\rm He}^+}&=&Y_{e^-},\label{eq-yhhe}
\end{eqnarray}
where $Y_p=0.7$ and $Y_\alpha=0.075$ are the numbers of protons and
$^4$He nuclei per nucleon, respectively ($Y_p+4Y_\alpha=1$ for 
consistency with a medium where the dominant nuclei are protons and 
$\alpha$ particles). 

Based on the chemical equilibrium for the reactions in Eqs.~(\ref{eq-eh}) 
and (\ref{eq-ehe}), we obtain
\begin{eqnarray}
\frac{Y_{\rm H}}{Y_{{\rm H}^{+}}}&=&\rho Y_{e^-}N_A
\left(\frac{g_{\rm H}}{4}\right)\left(\frac{2\pi\hbar^2}{m_ekT}\right)^{3/2}
\exp\left(\frac{I_{\rm H}}{kT}\right)=7.47\times 10^{-2}Y_{e^-},\label{eq-hh}\\
\frac{Y_{{\rm He}^+}}{Y_{\rm He}}&=&\frac{1}{\rho Y_{e^-}N_A}
\left(\frac{2g_{{\rm He}^+}}{g_{\rm He}}\right)
\left(\frac{m_ekT}{2\pi\hbar^2}\right)^{3/2}\exp\left(-\frac{I_{\rm He}}{kT}\right)
=\frac{1.52\times 10^{-2}}{Y_{e^-}},\label{eq-hehe}
\end{eqnarray}
where $m_e$ is the electron rest mass,
$g_{\rm H}\approx 4.8$, $g_{{\rm He}^+}\approx 2$, and 
$g_{\rm He}\approx 1$ are the partition
functions of the corresponding species, and $I_{\rm H}=13.6$~eV and 
$I_{\rm He}=24.6$~eV are the first ionization potentials for the ground states
of the H and He atoms, respectively. There is some subtlety in calculating
the partition functions as discussed in Appendix~\ref{sec-part}. However,
this does not affect the general results: nearly all of the H atoms are 
ionized and nearly all of the He atoms remain neutral. Solving 
Eqs.~(\ref{eq-yh})--(\ref{eq-hehe}), we obtain $Y_{e^-}\approx 0.67$,  
$Y_{{\rm H}^+}\approx 0.667$, $Y_{\rm H}\approx 0.033$, 
$Y_{\rm He}\approx 7.33\times 10^{-2}$, and 
$Y_{{\rm He}^+}\approx 1.7\times 10^{-3}$.

\begin{figure}
\includegraphics[scale=0.3]{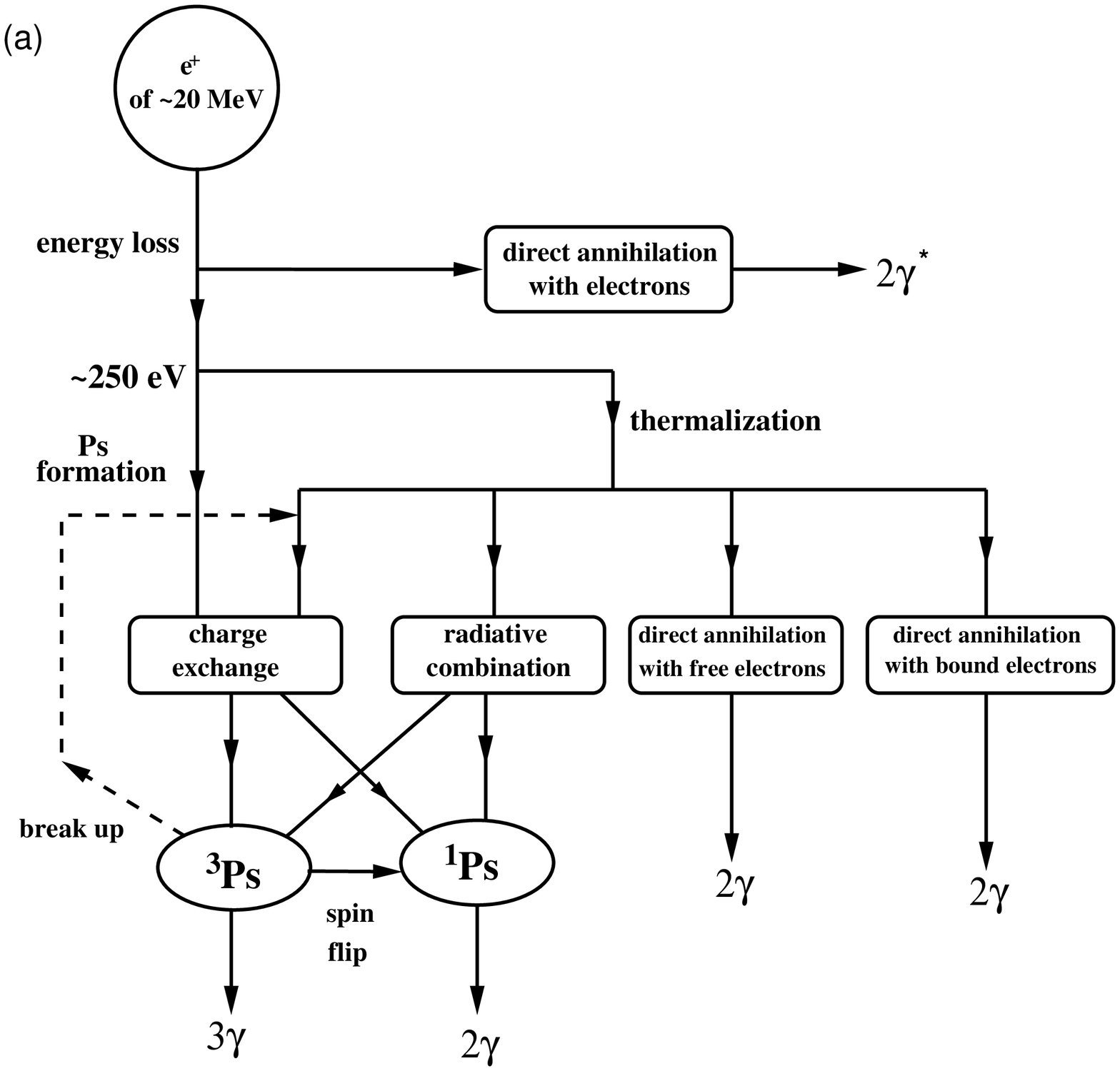}%

\includegraphics[scale=0.3]{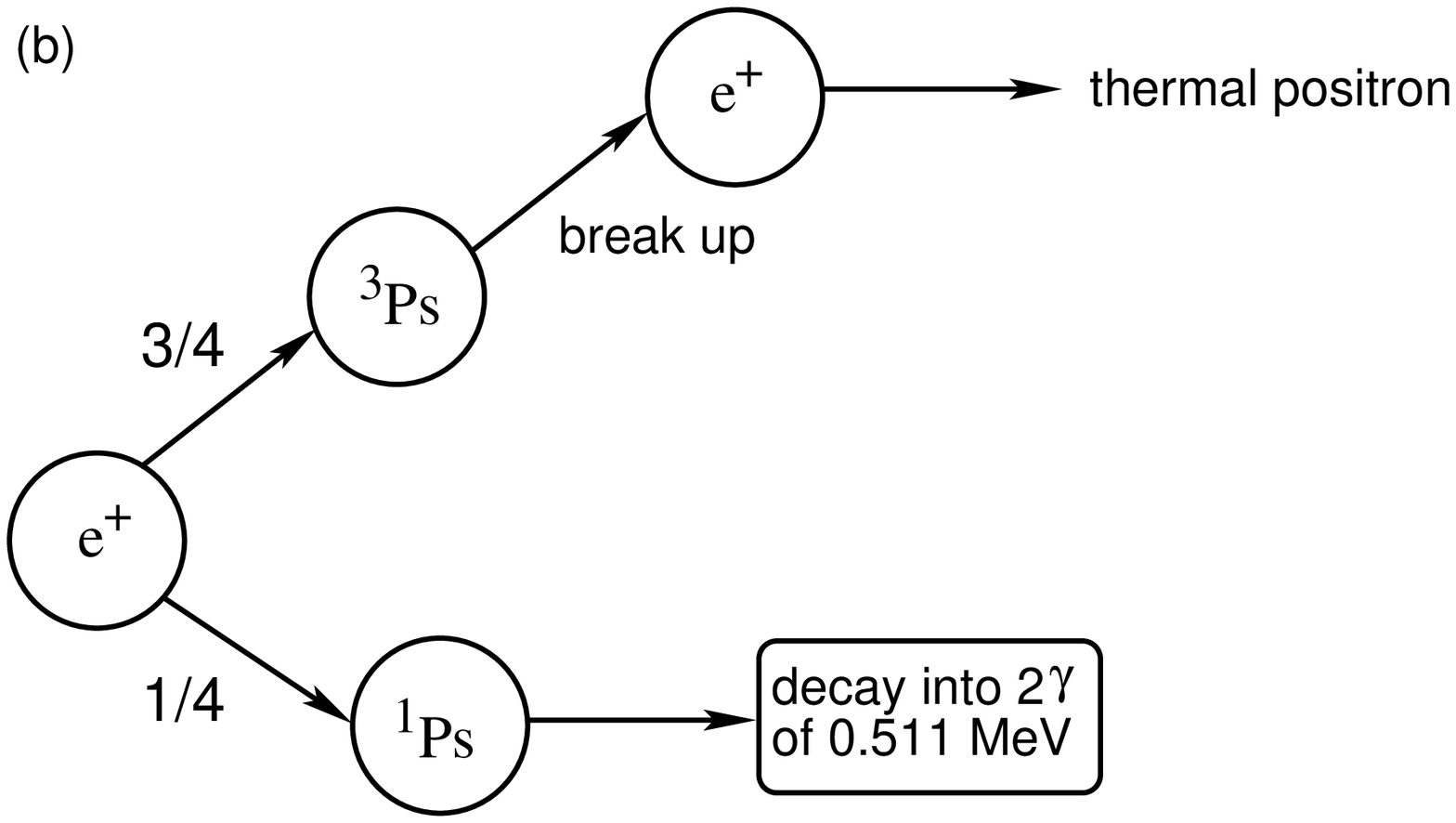}%
\includegraphics[scale=0.3]{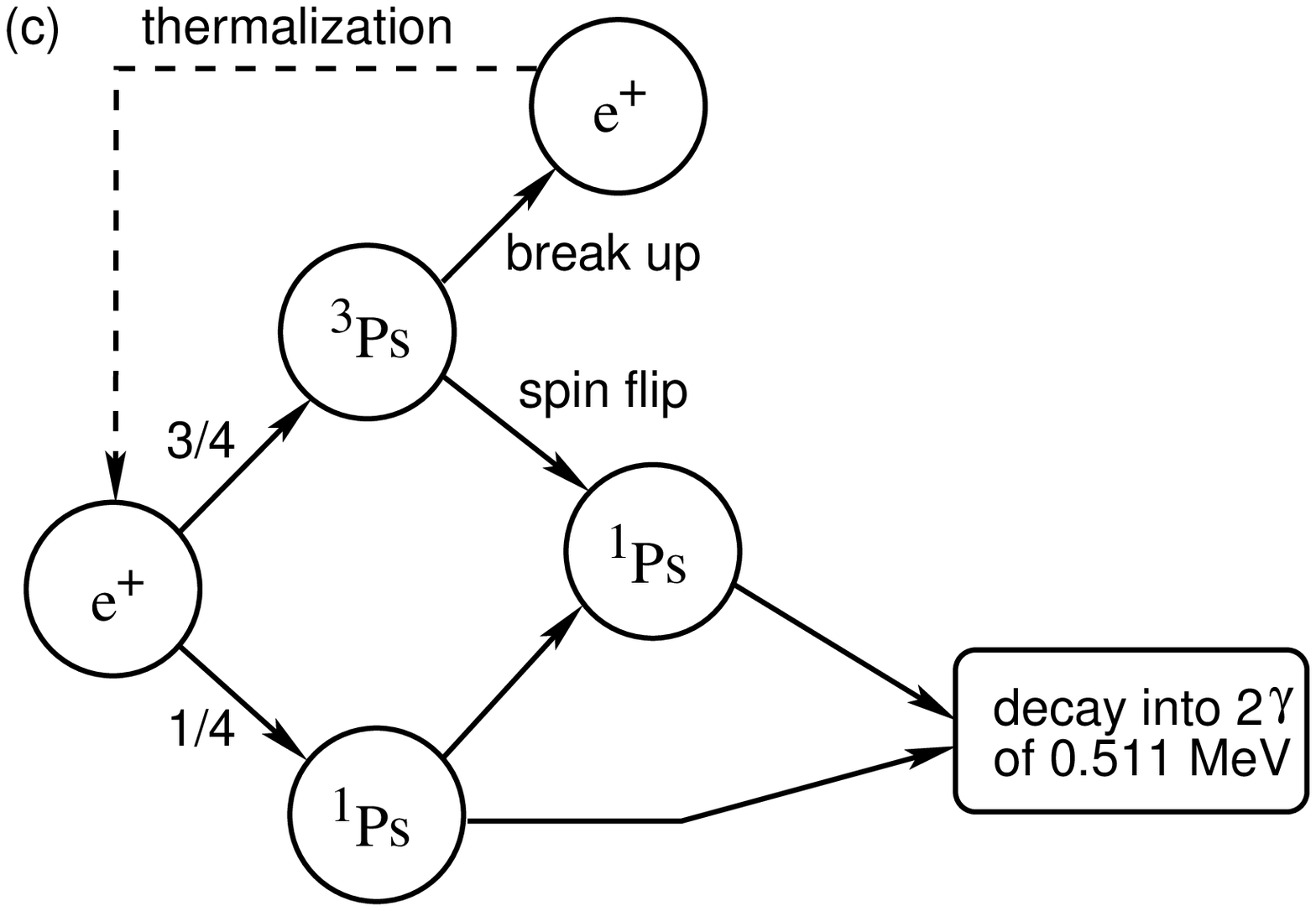}%
\caption{(a) Summary of the processes involved in the $\gamma$-ray emission
from annihilation of positrons with an initial energy of $\sim 20$~MeV.
Direct annihilation with electrons before thermalization produces $\gamma$
rays (indicated by $\gamma^*$) with energies shifted far from 0.511~MeV.
Positronium (Ps) formation with free and bound electrons is referred to as
radiative combination and charge exchange, respectively. (b) Summary of
the consequences of Ps formation that occurs after the positron {\it kinetic}
energy drops below $\sim 250$~eV but before thermalization. (c) Summary 
of the consequences of Ps formation after thermalization.
\label{fig-posa}}
\end{figure}

The positrons produced by $\bar\nu_e$ absorption on protons have
an average energy $\langle E_{e^+}\rangle\sim 
[F_5(\eta_{\bar\nu_e})/F_4(\eta_{\bar\nu_e})]T_{\bar\nu_e}\sim 21.2$~MeV.
As they pass through the essentially ionized plasma of the stellar surface
region, they can lose energy through many processes, directly
annihilate, and form Ps. The $^1$Ps formed decays into two photons
with an extremely short lifetime of $\tau_{2\gamma}=1.25\times 10^{-10}$~s 
while the $^3$Ps
decays into three photons
with a much longer lifetime of $\tau_{3\gamma}=1.42\times 10^{-7}$~s.
Consequently, once formed, $^1$Ps immediately decays while
$^3$Ps can be broken up or converted into $^1$Ps by the following
reactions before it decays:
\begin{eqnarray}
e^-+{^3{\rm Ps}}&\to&e^-+e^++e^-,\label{eq-psione}\\
{\rm H}+{^3{\rm Ps}}&\to&{\rm H}+e^++e^-,\\
{\rm H}^++{\rm Ps}&\to&{\rm H}+e^+,\label{eq-psionh}\\
{\rm He}^++{\rm Ps}&\to&{\rm He}+e^+,\label{eq-psionhe}\\
e^-+{^3{\rm Ps}}&\to&e^-+{^1{\rm Ps}},\\
{\rm H}+{^3{\rm Ps}}&\to&{\rm H}+{^1{\rm Ps}}.\label{eq-31h}
\end{eqnarray}
Note that the breakup reactions in Eqs.~(\ref{eq-psionh}) and 
(\ref{eq-psionhe}) are the reverse processes of the Ps formation 
reactions in Eqs.~(\ref{eq-psh}) and (\ref{eq-pshe}). Although all the 
processes in Eqs.~(\ref{eq-daf})--(\ref{eq-pshe}) and
(\ref{eq-psione})--(\ref{eq-31h}) are involved (to varying degrees)
in the $\gamma$-ray emission from positron annihilation 
(see Fig.~\ref{fig-posa}a),
for the adopted stellar conditions, the net result is 
rather simple: (1) Before thermalization, $\approx 12\%$ of the positrons 
directly annihilate and $\approx 4\%$ form Ps. 
Based on the available spin 
states, 1/4 of the Ps formed are $^1$Ps, which immediately
decay into two 0.511~MeV $\gamma$ rays with line widths of $\sim 6$~keV, 
and 3/4 of the Ps formed are $^3$Ps, which are immediately
broken up  (see Fig.~\ref{fig-posa}b). 
The positrons released by breakup of $^3$Ps are quickly
thermalized. (2) So effectively $\approx 87\%$ of the initial positrons are 
thermalized. Essentially all of the thermal positrons form Ps. 
The $^3$Ps formed are immediately broken up or
converted into $^1$Ps (see Fig.~\ref{fig-posa}c). 
The positrons released by breakup of $^3$Ps are 
again quickly thermalized. All the $^1$Ps formed after thermalization 
immediately decay into two 0.511~MeV $\gamma$ rays with line widths of 
$\sim 2$~keV.

\subsection{Thermalization of Positrons}
\label{sec-thpos}
As mentioned above, positrons can lose energy through many processes.
For the adopted stellar conditions, the dominant energy loss is due to excitation
of the free electrons in the plasma. The thermalization of positrons spans 
the highly relativistic and nonrelativistic regimes.  We use the general
results on positron interaction with plasma electrons given in 
Ref.~\cite{gould1}. In the relativistic regime, the energy
loss per unit length of propagation is
\begin{equation}
-\left(\frac{dE_{e^+}}{dx}\right)_{\rm ex,pl}=
4\pi\rho Y_{e^-}N_A\left(\frac{e^4}{m_ev^2}\right)B_{\rm rel}
=4.88\times 10^{-9}\left(\frac{c}{v}\right)^2Y_{e^-}B_{\rm rel}\ 
{\rm MeV\ cm}^{-1},
\label{eq-dedxrel}
\end{equation}
where
\begin{equation}
B_{\rm rel}=\ln\left[\frac{\sqrt{2\delta(\gamma-1)}\,m_evc}{\hbar\omega_p}\right]
+b(\gamma,\delta).
\end{equation}
In the above equations, $v$ is the positron velocity, $c$ is the speed of light,
$\gamma\equiv1/\sqrt{1-(v/c)^2}$, $\delta$ is the maximum fraction of the
positron energy lost in a single interaction and is taken to be 1/2, 
$\omega_p=\sqrt{4\pi\rho Y_{e^-}N_Ae^2/m_e}$ is the plasma frequency,
and
\begin{eqnarray}
b(\gamma,\delta)&\equiv&
-\left(\frac{\gamma^2-1}{2\gamma^2}\right)\delta
+\frac{1}{8}\left(\frac{\gamma-1}{\gamma}\right)^2\delta^2\nonumber\\
&&-\frac{1}{2}\left(\frac{\gamma-1}{\gamma+1}\right)\left[
\left(\frac{\gamma+2}{\gamma}\right)\delta
-\left(\frac{\gamma^2-1}{\gamma^2}\right)\delta^2
+\frac{1}{3}\left(\frac{\gamma-1}{\gamma}\right)^2\delta^3\right]\nonumber\\
&&+\frac{1}{2}\left(\frac{\gamma-1}{\gamma+1}\right)^2\left[
\left(\frac{1}{2}+\frac{1}{\gamma}+\frac{3}{2\gamma^2}\right)\frac{\delta^2}{2}
-\left(\frac{\gamma-1}{\gamma}\right)^2
\left(\frac{\delta^3}{3}-\frac{\delta^4}{4}\right)\right].
\label{eq-brel}
\end{eqnarray}
The energy loss rate in the nonrelativistic regime is obtained by replacing
$B_{\rm rel}$ in Eq.~(\ref{eq-dedxrel}) with
\begin{equation}
B_{\rm nr}=\ln\left[\frac{\sqrt{\delta}\,m_ev^2}{\hbar\omega_p}\right]-
\varepsilon-\Re\psi(i\alpha c/v),
\end{equation}
where $\varepsilon\approx0.577$ is Euler's constant, $\alpha=e^2/(\hbar c)$
is the fine-structure constant, and $\Re\psi(z)$ is
the real part of the digamma function $\psi(z)\equiv(d\Gamma/dz)/\Gamma(z)$
with $\Gamma(z)$ being the Gamma function. 

The positrons produced by $\bar\nu_e$ absorption on
protons have an average energy $\langle E_{e^+}\rangle\sim 21.2$~MeV 
corresponding to $\langle\gamma\rangle\sim 41.5$. Once thermalized, they have 
an average {\it kinetic} energy $E_{\rm th}=(3/2)kT=1.98$~eV corresponding to 
$\gamma_{\rm th}-1=3.87\times 10^{-6}$. Using the energy loss rate
discussed above, we can estimate the total distance $(\Delta x)_{\rm th}$
and time $(\Delta t)_{\rm th}$ covered by a typical positron during 
thermalization. We proceed by noting that both $-(dE_{e^+}/dx)_{\rm ex,pl}$ 
and $-(dE_{e^+}/dt)_{\rm ex,pl}=-v(dE_{e^+}/dx)_{\rm ex,pl}$ increase as 
the positron energy decreases. Consequently, the bulk of  $(\Delta x)_{\rm th}$
and  $(\Delta t)_{\rm th}$ is covered before positrons become nonrelativistic,
and we obtain
\begin{equation}
(\Delta x)_{\rm th}\sim\int_{\langle E_{e^+}\rangle}^{m_ec^2}
\frac{dE_{e^+}}{(dE_{e^+}/dx)_{\rm ex,pl}}\sim\frac{\langle E_{e^+}\rangle}
{4\pi\rho Y_{e^-}N_A[e^4/(m_ec^2)]B_{\rm rel}}\sim 3\times 10^8\ {\rm cm},
\end{equation}
where we have taken advantage of the slowly-varying function
$B_{\rm rel}$ in evaluating the integral and used $B_{\rm rel}\sim 20$
for the adopted stellar conditions. The distance $(\Delta x)_{\rm th}$
estimated above corresponds to a thermalization timescale
$(\Delta t)_{\rm th}\sim(\Delta x)_{\rm th}/c\sim 10^{-2}$~s.

The energy loss rate discussed above is shown as a function of 
$\gamma-1$ in Fig.~\ref{fig-dedx}. For comparison, the rates due to the other 
processes discussed in Appendix~\ref{sec-xpos} are also shown. It can be
seen that the energy loss due to excitation of the free electrons in the plasma
dominates over the entire energy range between production and
thermalization of positrons. We have numerically evaluated 
$(\Delta x)_{\rm th}$ and $(\Delta t)_{\rm th}$ first using the actual energy loss 
rate due to excitation of the free electrons in the plasma only and then
including the actual rates of the other processes discussed in 
Appendix~\ref{sec-xpos}. The inclusion of the other processes has rather
small effect and reduces
$(\Delta x)_{\rm th}$ from $2.98\times 10^8$ to $2.49\times 10^8$~cm and
$(\Delta t)_{\rm th}$ from $10^{-2}$ to $8.4\times 10^{-3}$~s.

\begin{figure}
\includegraphics[scale=0.25,angle=270]{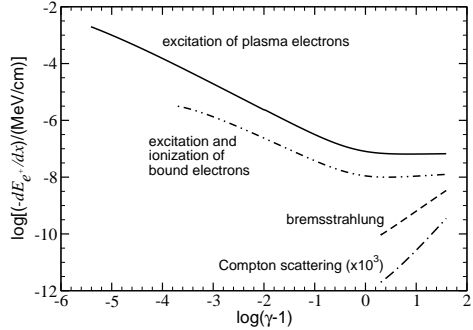}%
\caption{Rates of positron energy loss $-dE_{e^+}/dx$ as functions of 
$\gamma -1$ for excitation of free electrons in the plasma (solid curve),
excitation and ionization of bound electrons in atoms (dash-dot-dotted curve),
bremsstrahlung (dashed curve), and Compton scattering on thermal photons
(dashed-dotted curve). The rate for the first process is shown down to a positron
{\it kinetic} energy of 2~eV
(the thermal energy for the stellar conditions adopted here), 
that for the second process is shown down to a positron
{\it kinetic} energy of 100~eV (the average excitation 
energy being 15 and 41.5~eV for H and He atoms, respectivley), 
and those for the last two processes are
shown only for relativistic positrons with $\gamma\geq 3$. Note that the 
actual rate multiplied by a factor of $10^3$ is shown for Compton scattering.   
\label{fig-dedx}}
\end{figure}

\subsection{Direct Annihilation of Positrons before Thermalization}
\label{sec-anpos}

In addition to losing energy, a positron can directly annihilate with a free 
or bound electron before being thermalized. As the thermalization 
distance $(\Delta x)_{\rm th}$ is predominantly covered while
the positron is still relativistic, direct annihilation before
thermalization also predominantly occurs during the relativistic regime
(see below). 
We ignore the $\gamma$ rays produced from such annihilation as the
large positron velocity causes large shifts from the 0.511~MeV line emitted 
in the center-of-mass frame for the annihilating positron and electron. 
The quantity of interest to us is the probability for direct annihilation
of the positron before thermalization.

We first consider direct annihilation with
free electrons. For a fast positron ($v\gg 2\pi\alpha c$) annihilating with 
a free electron at rest, the cross section is
\begin{equation}
\sigma_{\rm da,f}^{\rm fast}=\left(\frac{e^2}{m_ec^2}\right)^2\frac{\pi}{\gamma+1}
\left[\frac{\gamma^2+4\gamma+1}{\gamma^2-1}\ln(\gamma+
\sqrt{\gamma^2-1})-\frac{\gamma+3}{\sqrt{\gamma^2-1}}\right].
\label{eq-xsdaf}
\end{equation}
The above result ignores the Coulomb interaction between the positron and
the electron. This interaction becomes important at lower positron velocities,
for which the cross section \cite{gould2} can be obtained 
by adding a ``Coulomb focusing'' factor to the low-velocity limit of the result in 
Eq.~(\ref{eq-xsdaf}):
\begin{equation}
\sigma_{\rm da,f}^{\rm slow}=\left(\frac{e^2}{m_ec^2}\right)^2
\frac{2\pi^2\alpha(c/v)^2}{1-\exp(-2\pi\alpha c/v)}.
\end{equation}
The probability for direct annihilation with free electrons before thermalization
can be estimated as
\begin{equation}
P_{\rm da,f}\sim Y_{e^-}\rho N_A\int_{\langle E_{e^+}\rangle}^{m_ec^2}
\frac{\sigma_{\rm da,f}^{\rm rel}\,dE_{e^+}}{(dE_{e^+}/dx)_{\rm ex,pl}}\sim
\frac{1}{4B_{\rm rel}}\int_1^{\langle\gamma\rangle}
\frac{\ln(2\gamma)}{\gamma}d\gamma\sim 0.1,\label{eq-pdaf}
\end{equation} 
where we have considered only the relativistic regime and used the excitation 
of free electrons as the dominant energy loss mechanism with 
$B_{\rm rel}\sim 20$. Note that although for low positron velocities
$\sigma_{\rm da,f}^{\rm slow}$ increases as $\sim 1/v^2$, this is cancelled by 
the same increase of the corresponding $-(dE_{e^+}/dx)_{\rm ex,pl}$
[see Eq.~(\ref{eq-dedxrel})]. Thus, direct annihilation with free electrons 
before thermalization predominantly occurs when the predominant part of
$(\Delta x)_{\rm th}$ is covered, i.e., when the positron is still relativistic
(see Sec.~\ref{sec-thpos}).
Note also that the probability of this occurrence depends 
logarithmically on the initial positron energy and the density of the medium 
(through $B_{\rm rel}$).

\begin{figure}
\includegraphics[scale=0.25,angle=270]{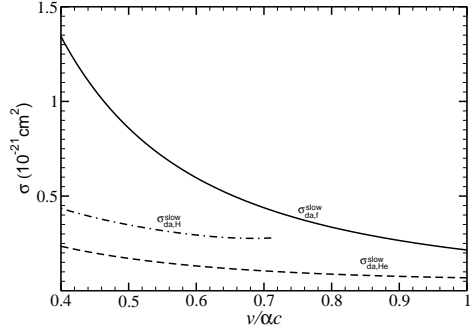}%
\caption{Cross sections for direct annihilation of slow positrons with
free electrons ($\sigma_{\rm da,f}^{\rm slow}$, solid curve) and with the
electrons in the H ($\sigma_{\rm da,H}^{\rm slow}$ \cite{bha}, dash-dotted 
curve) and He ($\sigma_{\rm da,He}^{\rm slow}$ \cite{hum}, dashed curve) 
atoms as functions of the positron velocity $v$ in units of $\alpha c$.
\label{fig-da}}
\end{figure}

The cross section $\sigma_{\rm da,b}^{\rm fast}$ for direct annihilation of 
a fast positron with a bound electron in an atom (or ion) is the same as 
$\sigma_{\rm da,f}^{\rm fast}$\,. In contrast, for $v\lesssim\alpha c$, 
\begin{equation}
\sigma_{\rm da,b}^{\rm slow}=\pi Z_{\rm eff}
\left(\frac{e^2}{m_ec^2}\right)^2\frac{c}{v}\,,
\end{equation}
where $Z_{\rm eff}$ is a function of $v$ and depends on the atom
(see Appendix~\ref{sec-apos} for the case of the H atom).
The cross sections $\sigma_{\rm da,H}^{\rm slow}$ and 
$\sigma_{\rm da,He}^{\rm slow}$ for direct annihilation with the
electrons in the H \cite{bha} and He \cite{hum} atoms, respectively, are 
compared with $\sigma_{\rm da,f}^{\rm slow}$ in Fig.~\ref{fig-da}.
As  $\sigma_{\rm da,f}^{\rm slow}>\sigma_{\rm da,H}^{\rm slow},\ 
\sigma_{\rm da,He}^{\rm slow}$ and $Y_{e^-}\gg Y_{\rm H},\ Y_{\rm He}$,
the probability for direct annihilation with bound electrons at low positron
velocities is much smaller than the corresponding probability in the case
of free electrons, which is already very small. On the other hand,
inclusion of the bound electrons increases the number of targets for 
direct annihilation in the relativistic regime. This gives
$P_{\rm da}\approx P_{\rm da,f}+P_{\rm da,H}+P_{\rm da,He}\approx
(Y_e/Y_{e^-})P_{\rm da,f}$, where
$P_{\rm da}$ is the total probability of direct annihilation before 
thermalization and $Y_e/Y_{e^-}=1.27$.
Using the actual cross sections and rates of various energy loss processes 
over the entire thermalization regime, we have numerically calculated the 
probability of direct 
annihilation with free electrons before thermalization and found 
$P_{\rm da,f}=9.52\%$. So the total probability including annihilation with
bound electrons is $P_{\rm da}\approx 12\%$.

\subsection{Ps Formation before Thermalization}
\label{sec-pspos}
\begin{figure}
\includegraphics[scale=0.25,angle=270]{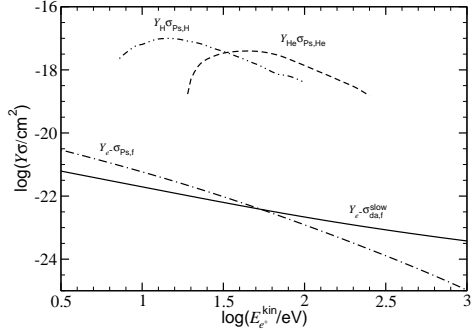}%
\caption{Cross section multiplied by the number
of targets per nucleon for direction annihilation of slow positrons
with free electrons ($Y_{e^-}\sigma_{\rm da,f}^{\rm slow}$, solid curve)
as a function of $E_{e^+}^{\rm kin}$ compared with the corresponding
quantities for
Ps formation with free electrons ($Y_{e^-}\sigma_{\rm Ps,f}$, 
dash-dottd curve) and with the electrons in the H 
($Y_{\rm H}\sigma_{\rm Ps,H}$, dash-dot-dotted curve) and 
He ($Y_{\rm He}\sigma_{\rm Ps,He}$, dashed curve) atoms.
The numbers of targets per nucleon used are
$Y_{e^-}=0.67$, $Y_{\rm H}=0.033$, and $Y_{\rm He}=7.33\times 10^{-2}$. 
The measured cross sections $\sigma_{\rm Ps,H}$ and 
$\sigma_{\rm Ps,He}$ for $E_{e^+}^{\rm kin}$ above the threshold
of 6.8 and 17.8~eV but below 100 and 250~eV are taken from 
Refs.~\cite{zhou} and \cite{overton}, respectively. 
The probabilities for Ps formation with the electrons 
in the H and He atoms are negligible for $E_{e^+}^{\rm kin}>100$ and 
250~eV, respectively, before thermalization as $\sigma_{\rm Ps,H}$ and 
$\sigma_{\rm Ps,He}$ rapidly decrease for such $E_{e^+}^{\rm kin}$.
\label{fig-xsps}}
\end{figure}

In addition to direct annihilation, positrons can form Ps with free 
and bound electrons before thermalization. As discussed above,
direct annihilation predominantly occurs in the relativistic regime.
As positrons become
more and more nonrelativistic, Ps formation becomes more and more
likely. The cross section for direct annihilation of slow positrons
with free electrons multiplied
by the number of free electrons per nucleon in the stellar surface
region under consideration, $Y_{e^-}\sigma_{\rm da,f}^{\rm slow}$, 
is compared in Fig.~\ref{fig-xsps} with the corresponding quantities 
$Y_{e^-}\sigma_{\rm Ps,f}$,
$Y_{\rm H}\sigma_{\rm Ps,H}$, and
$Y_{\rm He}\sigma_{\rm Ps,He}$ for Ps formation with free electrons
and the electrons in the H and He atoms, respectively. It can be seen 
that $Y_{\rm H}\sigma_{\rm Ps,H}$ and
$Y_{\rm He}\sigma_{\rm Ps,He}$ 
are extremely large at $6.8<E_{e^+}^{\rm kin}<100$~eV and 
$17.8<E_{e^+}^{\rm kin}<250$~eV, respectively.
As a result, once positrons are slowed down to 
$E_{e^+}^{\rm kin}\lesssim 250$~eV, they predominantly form Ps.
Using the cross sections for Ps formation $\sigma_{\rm Ps,f}$
(see Appendix~\ref{sec-apos}), $\sigma_{\rm Ps,H}$ \cite{zhou}, and
$\sigma_{\rm Ps,He}$ \cite{overton} 
and the rates of various energy loss processes, we find that the 
probability for Ps formation with the electrons in the H and He atoms
before thermalization is 
$P_{\rm Ps,H}\approx 1\%$ and $P_{\rm Ps,He}\approx 3\%$,
respectively, and that Ps formation with free electrons before 
thermalization is negligible (with a probability of $\ll 1\%$). So the
total probability for Ps formation before thermalization is
$P_{\rm Ps}\approx P_{\rm Ps,H}+P_{\rm Ps,He}\approx 4\%$.

Based on the available spin states, $^1$Ps and $^3$Ps constitute
1/4 and 3/4 of the Ps formed, respectively. Thus,
a fraction $P_{\rm Ps}/4\approx 1\%$ 
of the initial positrons form $^1$Ps before thermalization and
immediately decay into two $\gamma$ rays due to the extremely
short lifetime of $^1$Ps.
The velocity distribution of the decaying $^1$Ps gives rise to 
a spread in the energies of these $\gamma$ rays, which are centered 
at 0.511~MeV. This spread can be characterized by a formal full width 
at half maximum (FWHM) of $\sim 6$~keV \cite{murphy}.
In contrast, the relatively long lifetime of $^3$Ps allows it to be
broken up under the adopted stellar conditions. The dominant
breakup reactions are $e^-+{^3{\rm Ps}}\to e^-+e^++e^-$
and ${\rm H}^++{\rm Ps}\to{\rm H}+e^+$
with rates of $\approx 4.5\times 10^8$ and
$7.7\times 10^7$~s$^{-1}$, respectivley \cite{murphy}.
The total rate for breakup of $^3$Ps before thermalization  is 
$\lambda_{3\to 0}^{\rm bt}\approx 5.3\times 10^8$~s$^{-1}$,
to be compared with the rate of 
$\tau_{3\gamma}^{-1}=7.04\times 10^6$~s$^{-1}$ for $^3$Ps decay.
The rates of the above breakup reactions and many other reactions
involved in the $\gamma$-ray emission from positron annihilation
are given in Table~\ref{tab-rates} for the adopted stellar conditions.  
Thus, before thermalization a fraction $(3/4)P_{\rm Ps}\approx 3\%$ 
of the initial positrons form $^3$Ps, which are immediately broken up. 
The positrons released by the breakup 
reactions are quickly thermalized. As a result,
effectively a fraction $1-P_{\rm da}-(P_{\rm Ps}/4)\approx 87\%$ 
of the initial positrons are thermalized.

\begin{table}
\caption{Probabilities and rates of various processes involved in the
$\gamma$-ray emission from positron
annihilation in the stellar surface region. The conditions in this region are
characterized by $\rho=1.59\times 10^{-8}\ {\rm g}\ {\rm cm}^{-3}$,
$T=1.53\times 10^4$~K, $Y_{e^-}\approx 0.67$, $Y_{{\rm H}^+}\approx 0.667$,
$Y_{\rm H}\approx 0.033$, $Y_{\rm He}\approx 7.33\times 10^{-2}$, and
$Y_{{\rm He}^+}\approx 1.7\times 10^{-3}$. Note that the rates for breakup
and conversion of $^3$Ps depend on the formation mode of $^3$Ps.
The rates with the superscript ``bt'' correspond to the $^3$Ps 
formed before thermalization and those with the superscript ``Ps,H'' 
correspond to the $^3$Ps formed by the reaction 
$e^++{\rm H}\to{\rm Ps}+{\rm H}^+$ after thermalization.
\label{tab-rates}}
 \begin{ruledtabular}
\begin{tabular}{lll}
Processes&before thermalization&after thermalization\\\hline
$e^++e^-\to\gamma+\gamma$&$P_{\rm da,f}=9.52\%$&
$\lambda_{\rm da,f}=7.7\times 10^2\ {\rm s}^{-1}$\\
$e^++{\rm H}\to{\rm H}^++\gamma+\gamma$&$P_{\rm da,H}\approx 0.47\%$&
$\lambda_{\rm da,H}\approx 13\ {\rm s}^{-1}$\\
$e^++{\rm He}\to{\rm He}^++\gamma+\gamma$&$P_{\rm da,He}\approx 2.08\%$&
$\lambda_{\rm da,He}\approx 15\ {\rm s}^{-1}$\\
$e^++e^-\to{\rm Ps}+\gamma$&$P_{\rm Ps,f}\ll 1\%$&
$\lambda_{\rm Ps,f}=4.6\times 10^3\ {\rm s}^{-1}$\\
$e^++{\rm H}\to{\rm Ps}+{\rm H}^+$&$P_{\rm Ps,H}\approx 1\%$&
$\lambda_{\rm Ps,H}\approx 9.9\times 10^4\ {\rm s}^{-1}$\\
$e^++{\rm He}\to{\rm Ps}+{\rm He}^+$&$P_{\rm Ps,He}\approx 3\%$&
$\lambda_{\rm Ps,He}\approx 4.6\ {\rm s}^{-1}$\\
$e^-+{^3{\rm Ps}}\to e^-+e^++e^-$&
$\lambda_{3\to 0,{\rm f}}^{\rm bt}\approx 4.5\times 10^8\ {\rm s}^{-1}$&
$\lambda_{3\to 0,{\rm f}}^{\rm Ps,H}\approx 1.2\times 10^7\ {\rm s}^{-1}$\\
${\rm H}+{^3{\rm Ps}}\to{\rm H}+e^++e^-$&
$\lambda_{3\to 0,{\rm H}}^{\rm bt}\approx 7.9\times 10^6\ {\rm s}^{-1}$&
$\lambda_{3\to 0,{\rm H}}^{\rm Ps,H}\approx 1.1\times 10^5\ {\rm s}^{-1}$\\
${\rm H}^++{\rm Ps}\to{\rm H}+e^+$&
$\lambda_{3\to 0,{\rm H}^+}^{\rm bt}\approx 7.7\times 10^7\ {\rm s}^{-1}$&
$\lambda_{3\to 0,{\rm H}^+}^{\rm Ps,H}\approx 3.2\times 10^8\ {\rm s}^{-1}$\\
${\rm He}^++{\rm Ps}\to{\rm He}+e^+$&
$\lambda_{3\to 0,{\rm He^+}}^{\rm bt}\approx 9.8\times 10^4\ {\rm s}^{-1}$&
$\lambda_{3\to 0,{\rm He^+}}^{\rm Ps,He}\approx 5.5\times 10^4\ {\rm s}^{-1}$\\
$e^-+{^3{\rm Ps}}\to e^-+{^1{\rm Ps}}$&
$\lambda_{3\to 1,{\rm f}}^{\rm bt}\approx 1.9\times 10^6\ {\rm s}^{-1}$&
$\lambda_{3\to 1,{\rm f}}^{\rm Ps,H}\approx 5.1\times 10^8\ {\rm s}^{-1}$\\
${\rm H}+{^3{\rm Ps}}\to{\rm H}+{^1{\rm Ps}}$&
$\lambda_{3\to 1,{\rm H}}^{\rm bt}\approx 3.9\times 10^6\ {\rm s}^{-1}$&
$\lambda_{3\to 1,{\rm H}}^{\rm Ps,H}\approx 9.0\times 10^6\ {\rm s}^{-1}$\\
\end{tabular}
\end{ruledtabular}
\end{table}

\subsection{Fluxes of 0.511~MeV $\gamma$ Rays}
\label{sec-gflux}
The thermal positrons again can directly annihilate with or form Ps with
free or bound electrons. The rates for these processes are given in
Table~\ref{tab-rates} for the adopted stellar conditions. The dominant
reaction is Ps formation with the electron in the H atom with a rate of
$\lambda_{\rm Ps,H}=9.9\times 10^4$~s$^{-1}$ \cite{murphy}. 
The $^3$Ps formed is quickly converted
into $^1$Ps or broken up. The dominant conversion reaction is 
$e^-+{^3{\rm Ps}}\to e^-+{^1{\rm Ps}}$ with a rate of
$\lambda_{3\to 1}^{\rm Ps,H}=5.1\times 10^8$~s$^{-1}$, and
the dominant breakup reaction is
${\rm H}^++{\rm Ps}\to {\rm H}+e^+$ with a rate of
$\lambda_{3\to 0}^{\rm Ps,H}=3.2\times 10^8$~s$^{-1}$
\cite{murphy} (see also Table~\ref{tab-rates}). Whether directly formed
or produced by the conversion reaction, all the $^1$Ps  
immediately decay into two $\gamma$ rays
centered at 0.511~MeV with a FWHM of $\sim 2$~keV \cite{murphy}.
The positrons released
by the breakup reaction are again quickly thermalized and follow the 
same fate of thermal positrons as outlined above.

Based on the discussion in the preceding subsections, a fraction
$P_{\rm da}\approx 12\%$ of the positrons produced by $\bar\nu_e$ 
absorption on protons directly annihilate and a fraction
$P_{\rm Ps}\approx 4\%$ form Ps before being thermalized, and
the rest are thermalized on a timescale of $\sim 10^{-2}$~s, which is
much shorter than the timescale of $\tau=3$~s
governing the production of the initial positrons. The positrons
released by the breakup of $^3$Ps are also quickly thermalized.
As the positrons are produced approximately isotropically, their spatial 
distribution can be taken as uniform before and after thermalization. 
So in the region of interest to $\gamma$-ray emission, the time evolution 
for the numbers of various species per nucleon is governed by
\begin{eqnarray}
\frac{dY_{^1{\rm Ps,bt}}}{dt}&\approx&\frac{1}{4}P_{\rm Ps}
\lambda_{\bar\nu_ep}(t)-\frac{Y_{^1{\rm Ps,bt}}}{\tau_{2\gamma}},\\
\frac{dY_{^3{\rm Ps,bt}}}{dt}&\approx&\frac{3}{4}P_{\rm Ps}
\lambda_{\bar\nu_ep}(t)-\lambda_{3\to 0}^{\rm bt}Y_{^3{\rm Ps,bt}},\\
\frac{dY_{e^+}}{dt}&\approx&(1-P_{\rm da}-P_{\rm Ps})
\lambda_{\bar\nu_ep}(t)+\lambda_{3\to 0}^{\rm bt}Y_{^3{\rm Ps,bt}}
+\lambda_{3\to 0}^{\rm Ps,H}Y_{^3{\rm Ps}}
-\lambda_{\rm Ps,H}Y_{e^+},\\
\frac{dY_{^1{\rm Ps}}}{dt}&\approx&\frac{1}{4}\lambda_{\rm Ps,H}Y_{e^+}
+\lambda_{3\to 1}^{\rm Ps,H}Y_{^3{\rm Ps}}-
\frac{Y_{^1{\rm Ps}}}{\tau_{2\gamma}},\\
\frac{dY_{^3{\rm Ps}}}{dt}&\approx&\frac{3}{4}\lambda_{\rm Ps,H}Y_{e^+}
-(\lambda_{3\to 1}^{\rm Ps,H}+\lambda_{3\to 0}^{\rm Ps,H})Y_{^3{\rm Ps}},
\end{eqnarray}
where $Y_{^1{\rm Ps,bt}}$, $Y_{^3{\rm Ps,bt}}$, $Y_{^1{\rm Ps}}$, and
$Y_{^3{\rm Ps}}$ correspond to the numbers per nucleon for the
$^1$Ps and $^3$Ps formed before and after thermalization, and
$Y_{e^+}$ is the number of thermal positrons per nucleon. As the
timescale of $\tau=3$~s governing $\lambda_{\bar\nu_ep}(t)$ is much
longer than $\tau_{2\gamma}$,
$(\lambda_{3\to 0}^{\rm bt})^{-1}$, $(\lambda_{\rm Ps,H})^{-1}$, and
$(\lambda_{3\to 1}^{\rm Ps,H}+\lambda_{3\to 0}^{\rm Ps,H})^{-1}$,
the above differential equations can be solved to good approximation
by setting all the time derivatives to zero. Thus, we have
\begin{eqnarray}
Y_{^1{\rm Ps,bt}}(t)&\approx&\frac{1}{4}P_{\rm Ps}
\lambda_{\bar\nu_ep}(t)\tau_{2\gamma},\\
Y_{^1{\rm Ps}}(t)&\approx&\left(1-P_{\rm da}-\frac{1}{4}P_{\rm Ps}\right)
\lambda_{\bar\nu_ep}(t)\tau_{2\gamma}.
\end{eqnarray}
These results are equivalent to what is stated at the beginning of
Sec.~\ref{sec-pos} and in Sec.~\ref{sec-pspos}:
a fraction $P_{\rm Ps}/4\approx 1\%$ of the 
initial positrons form $^1$Ps before thermalization and effectively a fraction
$1-P_{\rm da}-(P_{\rm Ps}/4)\approx 87\%$ of them form $^1$Ps 
after thermalization.

Using a similar prescription to that for estimating the flux due to neutron
capture on protons, we find that at radius $r>R$
the flux of 0.511~MeV $\gamma$ rays  
from decay of the $^1$Ps formed before thermalization is
\begin{eqnarray}
\Phi_{\gamma(e^\pm),{\rm bt}}(r,t_r)&\approx&
\frac{Y_{^1{\rm Ps,bt}}(t)}{4\tau_{2\gamma}}\left(\frac{R}{r}\right)^2\rho N_A
\int_0^d\exp\left(-\frac{d-x}{l_{\gamma(e^\pm)e}}\right)dx,\label{eq-fpost}\\
&\approx&\frac{P_{\rm Ps}\lambda_{\bar\nu_ep}(t)}{16}\left(\frac{R}{r}\right)^2
\rho N_Al_{\gamma(e^\pm)e},\\
&\approx&\frac{P_{\rm Ps}E_B}{384\pi r^2\langle E_{\bar\nu_e}\rangle\tau}
\left(\frac{Y_p\langle\sigma_{\bar\nu_ep}\rangle}{Y_e\sigma_{\gamma(e^\pm)e}}\right)
\exp(-t/\tau),\\
&\approx&7.8\times10^{-7}\left(\frac{1\ {\rm kpc}}{r}\right)^2\exp\left(-t/\tau\right)
\ {\rm cm}^{-2}\ {\rm s}^{-1},
\end{eqnarray}
where the factor 1/4 in Eq.~(\ref{eq-fpost})
comes from the integration of the local flux at a point in
the stellar surface region over the forward solid angle. 
Note that although two 0.511~MeV 
$\gamma$ rays are emitted in $^1$Ps decay, only one contributes to the
emergent flux (the other being emitted towards the stellar interior). Likewise,
at radius $r>R$ the flux of 0.511~MeV $\gamma$ rays  
from decay of the $^1$Ps formed after thermalization is
\begin{eqnarray}
\Phi_{\gamma(e^\pm)}(r,t_r)&\approx&
\frac{[1-P_{\rm da}-(P_{\rm Ps}/4)]E_B}{96\pi r^2\langle E_{\bar\nu_e}\rangle\tau}
\left(\frac{Y_p\langle\sigma_{\bar\nu_ep}\rangle}{Y_e\sigma_{\gamma(e^\pm)e}}\right)
\exp(-t/\tau),\\
&\approx&6.8\times10^{-5}\left(\frac{1\ {\rm kpc}}{r}\right)^2\exp\left(-t/\tau\right)
\ {\rm cm}^{-2}\ {\rm s}^{-1}.
\end{eqnarray}
The $\gamma$ rays associated with 
$\Phi_{\gamma(e^\pm),{\rm bt}}(r,t_r)$ and $\Phi_{\gamma(e^\pm)}(r,t_r)$
differ in that the former have a FWHM of $\sim 6$~keV while the latter have
a FWHM of $\sim 2$~keV \cite{murphy}. 
Note that $\Phi_{\gamma(e^\pm),{\rm bt}}(r,t_r)$
has only a weak dependence on the density of the stellar surface region
through $P_{\rm Ps}$, which in turn depends logarithmically on the density
through $B_{\rm nr}$ associated with the energy loss due to excitation of 
free electrons [see similar density dependence for $P_{\rm da,f}$ through 
$B_{\rm rel}$ as exhibited in Eq.~(\ref{eq-pdaf})]. In contrast, apart from
the weak density dependence of $P_{\rm da}$ and $P_{\rm Ps}$, 
$\Phi_{\gamma(e^\pm)}(r,t_r)$ is sensitive to the breakup of
$^3$Ps and the conversion of 
$^3$Ps into $^1$Ps, the rates of which are proportional to the density
of the stellar surface region. For reasonable stellar conditions, 
$\Phi_{\gamma(e^\pm),{\rm bt}}(r,t_r)$ is overwhelmed by 
$\Phi_{\gamma(e^\pm)}(r,t_r)$.

\section{Discussion and Conclusions}
\label{sec-dis}
We have calculated the expected fluxes of 2.22 and 0.511~MeV $\gamma$ rays 
from neutron capture on protons and positron annihilation, respectively, 
following $\bar\nu_e$ absorption on protons in
the hydrogen envelope of an $11\,M_\odot$ star that undergoes core collapse
to produce a supernova. The $\gamma$-ray flux from neutron capture 
on protons exponentially 
decays on a timescale of $\tau_{\rm eff}=564$~s, which is determined by neutron 
decay and capture on protons and $^3$He nuclei. The peak flux is
$2.38\times 10^{-7}\ {\rm cm}^{-2}\ {\rm s}^{-1}$ for a supernova at a distance
of 1~kpc. In contrast, the $\gamma$-ray flux from positron annihilation follows the 
time evolution of the $\bar\nu_e$ luminosity. Although exponential decay on
a timescale of $\tau=3$~s is assumed here, the identical time evolution
for the $\bar\nu_e$ luminosity and the $\gamma$-ray flux from positron annihilation 
holds in general so long as the timescales for thermalization of positrons and Ps 
formation are much shorter than $\sim 1$~s. The peak flux in this case is
$6.8\times10^{-5}\ {\rm cm}^{-2}\ {\rm s}^{-1}$ for a supernova at a distance
of 1~kpc. Detection of the $\gamma$-ray fluxes quoted above 
is beyond the capability of 
current instruments, and perhaps even those planned for the near future.
For example, the proposed Advanced Compton Telescope \cite{act}
has a spectral resolution of 0.2--1\% over the energy range of
0.2--10~MeV and an angular resolution of $\sim 1^\circ$, which are ideal
for detecting the narrow $\gamma$-ray lines discussed here.
This instrument has a projected sensitivity of 
$5\times 10^{-7}\ {\rm cm}^{-2}\ {\rm s}^{-1}$ for narrow lines but an
exposure time of $10^6$~s is needed. As the fluxes of 2.22 and 0.511~MeV 
$\gamma$ rays discussed here only last for $\sim 10^3$ and $10$~s, respectively,
their detection requires much more sensitive instruments.

If the $\gamma$-ray fluxes discussed here can be detected,
they not only constitute a new kind of signals that
occur during the gap of several hours between the neutrino signals and the 
optical display of a supernova,
but may also provide a probe of the conditions in the surface layers of the
supernova progenitor. For example, both the peak and the decay timescale
$\tau_{\rm eff}$ of the $\gamma$-ray flux from neutron capture on protons
depend on the density of the stellar surface region. A higher density  
decreases the timescales for neutron capture on protons 
($\tau_{np}$) and $^3$He nuclei ($\tau_{n3}$), which
increases the peak flux and decreases $\tau_{\rm eff}$. On the other hand,
a higher $^3$He abundance decreases $\tau_{n3}$, and hence $\tau_{\rm eff}$,
but does not affect the peak flux as neutron capture on $^3$He nuclei consumes
neutrons without producing any $\gamma$ ray. As another example, 
the rates for the breakup of $^3$Ps and the conversion
of $^3$Ps into $^1$Ps are proportional to the density of the stellar surface region. 
While the $\gamma$-ray flux from positron annihilation quoted above
also applies approximately
to higher densities than adopted here, for sufficiently smaller densities
the $^3$Ps formed would predominantly decay into three $\gamma$ rays with 
a continuous spectrum instead of being broken up or converted. This would 
decrease the flux of 0.511~MeV
$\gamma$ rays from $^1$Ps decay. In the limit where no $^3$Ps are broken up
or converted, this flux is reduced to 1/4 of the value estimated here.
The above examples clearly illustrate
that neutrino-induced $\gamma$-ray emission from the
hydrogen envelope of a core-collapse supernova may serve as a useful 
probe of the conditions in the surface layers of the supernova progenitor.

\begin{acknowledgments}
We thank Wick Haxton for motivating us to carry out a detailed investigation of
neutrino-induced $\gamma$-ray emission from supernovae, John Beacom
for calling our attention to Ref.~\cite{rya}, and Icko Iben for providing the
electronic file of
the stellar model in Ref.~\cite{rito}. We also thank Robert Gould and
Richard Lingenfelter for helpful discussions. This work was supported in part by 
DOE grant DE-FG02-87ER40328.
\end{acknowledgments}

\appendix

\section{Cross Sections for $\bar\nu_e$ Absorption on Protons and Compton 
Scattering}
\label{sec-xs}
In this appendix we set $\hbar=c=1$.
The cross section for the reaction $\bar\nu_e+p\to n+e^+$ is \cite{vogel}
\begin{eqnarray}
\sigma_{\bar\nu_ep}&=&\frac{G_F^2\cos^2\theta_C}{\pi}(f^2+3g^2)(1+\delta_R)
(E_{\bar\nu_e}-\Delta)^2\left\{1-\frac{2[f^2+2(f+f_2)g+5g^2]}{f^2+3g^2}
\left(\frac{E_{\bar\nu_e}}{M_N}\right)\right\}\nonumber\\
&=&9.56\times 10^{-44}\left(\frac{E_{\bar\nu_e}-\Delta}{\rm MeV}\right)^2
\left[1-7.2\left(\frac{E_{\bar\nu_e}}{M_N}\right)\right]\ {\rm cm}^2,
\label{eq-nux}
\end{eqnarray}
where $G_F$ is the Fermi constant, $\theta_C$ is the Cabbibo angle
with $\cos\theta_C=0.9738$, $f=1$ and $g=1.27$ are the vector and 
axial vector coupling constants,
$f_2=3.706$ is the anomalous nucleon isovector magnetic moment,
$\delta_R\approx 0.024$ is the inner radiative corrections, 
$\Delta=M_n-M_p=1.293$~MeV is the difference between the neutron and proton
masses $M_n$ and $M_p$, and $M_N=(M_n+M_p)/2=938.9$~MeV.

The cross section for Compton scattering is
\begin{equation}
\sigma_{\gamma e}=\frac{\pi}{\epsilon_\gamma}\left(\frac{e^2}{m_e}\right)^2
\left[\left(1-\frac{2}{\epsilon_\gamma}-\frac{2}{\epsilon_\gamma^2}\right)
\ln(1+2\epsilon_\gamma)+\frac{1}{2}+\frac{4}{\epsilon_\gamma}
-\frac{1}{2(1+2\epsilon_\gamma)^2}\right],
\end{equation}
where $e$ is the magnitude of the electron charge, $m_e$ is the electron rest 
mass, and $\epsilon_\gamma=E_\gamma/m_e$ is the photon energy in units
of $m_e$.

\section{Partition Functions of Atoms}
\label{sec-part}
The number of electron energy states in an isolated atom is infinite. This
presents a problem in summing over these states to obtain the partition
function of the atom as the sum formally diverges. However, the application
of the partition function is sensible only when there are a large number of 
atoms. Consequently, the largest orbital radius of the electron in an atom
is physically restricted to the interatomic distance. For our problem, the
maximum radius $r_{\rm max}$ can be estimated from
\begin{equation}
\frac{4\pi}{3}r_{\rm max}^3\rho(Y_p+Y_\alpha)N_A=1.
\end{equation}
Take the H atom as an example. The largest orbital radius of the electron
is related to the maximum principal quantum number $n_{\rm max}$ as
$r_{\rm max}\sim n_{\rm max}^2\hbar^2/(m_ee^2)$. For our adopted stellar 
conditions, $n_{\rm max}\sim 25$. The partition function of the H atom is then
\begin{equation}
g_{\rm H}=4\sum_{n=1}^{n_{\rm max}}n^2\exp\left[-\frac{I_{\rm H}}{kT}
\left(1-\frac{1}{n^2}\right)\right]\approx 4.8,
\end{equation}
where the factor of 4 comes from the spin states for the proton
and the electron, and the factor $n^2$ accounts for the degeneracy of
the orbital states of the $n$th energy level. 

It can be seen that the partition function of the H atom under the stellar 
conditions of interest is dominated by the contribution from the ground state. 
This is also true for the He$^+$ ion and the He atom. As these two species
are minor components in the stellar region of interest, we take their partition 
functions to be given approximately by the contributions from the
corresponding ground states 
only, i.e., $g_{{\rm He}^+}\approx 2$ and $g_{\rm He}\approx 1$.

\section{Rates of Positron Energy Loss}
\label{sec-xpos}
In addition to excitation of the free electrons in the plasma, positrons can
also lose energy through excitation and ionization of the bound electrons in 
atoms and ions. Using the results from Ref.~\cite{gould3}, we find that 
the energy loss rate per unit length of propagation for the latter process is
\begin{eqnarray}
-\left(\frac{dE_{e^+}}{dx}\right)_{\rm ex,at}&\approx&
4\pi\rho N_A\left(\frac{e^4}{m_ev^2}\right)
(Y_{\rm H}B_{\rm H}+2Y_{\rm He}B_{\rm He})\nonumber\\
&=&4.88\times 10^{-9}\left(\frac{c}{v}\right)^2
(Y_{\rm H}B_{\rm H}+2Y_{\rm He}B_{\rm He})\ 
{\rm MeV\ cm}^{-1},
\label{eq-dedxa}
\end{eqnarray}
where $B_{\rm H}$ and $B_{\rm He}$ are of the form
\begin{equation}
B=\ln\left[\frac{\gamma\sqrt{2\delta(\gamma-1)}\,m_evc}{\overline{\Delta E}}\right]
-\frac{1}{2}\left(\frac{v}{c}\right)^2+b(\gamma,\delta).
\end{equation}
In the above equation, $\overline{\Delta E}$ is the average excitation energy
and $\overline{\Delta E}=15$ and 41.5~eV for H and He
atoms, respectively. In Eq.~(\ref{eq-dedxa}) we have ignored the contributions 
from the He$^+$ ions
as their abundance is much smaller than the abundances of H and He atoms.
Note that the energy loss due to excitation and ionization of the bound electrons
in atoms and ions is significant only when the positron {\it kinetic} energy is 
$E_{e^+}^{\rm kin}\gg\overline{\Delta E}$.

In the relativistic regime, three additional processes may be considered for
positron energy loss. We follow the discussion in Ref.~\cite{gould4} and first 
consider bremsstrahlung. In general the energy loss rate in an ionized plasma
differs from that in neutral matter. However, Ref.~\cite{gould4} showed that
the rates are the same for these two cases for $\gamma\lesssim 10^2$.
The relevant energy loss rate for our problem is
\begin{eqnarray}
-\left(\frac{dE_{e^+}}{dx}\right)_{\rm brem}&=&
4\alpha\left(\frac{e^2}{m_ec^2}\right)^2
E_{e^+}(2Y_p+6Y_\alpha)\left[\ln(2\gamma)-\frac{1}{3}\right]\nonumber\\
&=&2.10\times 10^{-11}\gamma\left[\ln(2\gamma)-\frac{1}{3}\right]\ 
{\rm MeV\ cm}^{-1}.
\end{eqnarray}
Positrons can also lose energy through Compton scattering on the photons
in the radiation field of the stellar surface region. The corresponding energy
loss rate is
\begin{equation}
-\left(\frac{dE_{e^+}}{dx}\right)_{\rm Comp}=\frac{32\pi}{9}
\left(\frac{e^2}{m_ec^2}\right)^2a_{\rm rad}T^4\gamma^2=2.30\times 10^{-16}
\gamma^2\ {\rm MeV\ cm}^{-1},
\end{equation}
where $a_{\rm rad}$ is the radiation constant. In the presence of a magnetic
field $\cal{B}$, positrons can lose energy through synchrotron radiation. This
process is similar to Compton scattering in that $\cal{B}$ can be viewed as a 
source of virtual photons. The relative importance of these two processes
can be gauged by comparing the energy densities in the magnetic and the
radiation fields:
\begin{equation}
\frac{{\cal{B}}^2/(8\pi)}{a_{\rm rad}T^4}=0.959\left(\frac{\cal{B}}{100\ {\rm G}}\right)^2.
\end{equation}

The energy loss rates $-(dE_{e^+}/dx)_{\rm ex,pl}$, $-(dE_{e^+}/dx)_{\rm ex,at}$,
$-(dE_{e^+}/dx)_{\rm brem}$, and $-(dE_{e^+}/dx)_{\rm Comp}$ are compared in
Fig.~\ref{fig-dedx}. Note that the last two rates increase with increasing 
$\gamma$ (at least for the positron energies of interest here) whereas the first two
rates decrease with increasing $\gamma$.
Therefore, if the positron energy loss through bremsstrahlung and Compton 
scattering is unimportant in the relativistic regime,
it can also be ignored in the nonrelativistic regime. Note also that 
$-(dE_{e^+}/dx)_{\rm ex,pl}$ exceeds $-(dE_{e^+}/dx)_{\rm Comp}$ by
a factor of at least $\sim 2\times 10^5$. So the positron energy loss through
synchrotron radiation can be ignored for ${\cal{B}}\ll 4\times 10^4$~G. 
We assume that ${\cal{B}}\ll 4\times 10^4$~G in the surface region of 
the star under consideration.

\section{Cross Sections for Direct Annihilation of Positrons
and Positronium Formation}
\label{sec-apos}
As an example of direct annihilation with bound electrons, we give 
the cross section $\sigma_{\rm da,H}^{\rm slow}$ for annihilation of a slow
positron with the electron in the H atom \cite{bha}:
\begin{equation}
\sigma_{\rm da,H}^{\rm slow}=\pi Z_{\rm eff,H}
\left(\frac{e^2}{m_ec^2}\right)^2\frac{c}{v},
\end{equation}
where $Z_{\rm eff,H}$ is a function of the positron velocity $v$ 
and can be approximated  as
\begin{eqnarray}
Z_{\rm eff,H}&\approx&8.868-7.838\left(\frac{v}{\alpha c}\right)
-102.77\left(\frac{v}{\alpha c}\right)^2+527.38\left(\frac{v}{\alpha c}\right)^3
\nonumber\\
&&-978.68\left(\frac{v}{\alpha c}\right)^4+773.15\left(\frac{v}{\alpha c}\right)^5
-197.17\left(\frac{v}{\alpha c}\right)^6.
\end{eqnarray}
The above approximation of $Z_{\rm eff,H}$ is valid for $v\lesssim 0.7\alpha c$, 
which corresponds to positron {\it kinetic} energy of 
$E_{e^+}^{\rm kin}<I_{\rm H}/2=6.8$~eV, i.e., below the threshold
for Ps formation with the electron in the H atom.
The cross sections $\sigma_{\rm da,f}^{\rm slow}$, 
$\sigma_{\rm da,H}^{\rm slow}$,  and $\sigma_{\rm da,He}^{\rm slow}$ 
\cite{hum} are compared in Fig.~\ref{fig-da}.

The relative importance of direct annihilation and Ps
formation depends on $E_{e^+}^{\rm kin}$. 
Direct annihilation is dominant in the relativistic regime.
As the positron becomes more and more nonrelativistic, Ps 
formation becomes more and more important. The cross section for a
nonrelativistic positron to form a Ps with a free electron at rest is
(see e.g., \cite{gould2,gould5})
\begin{equation}
\sigma_{\rm Ps,f}=\frac{2^7\pi\alpha}{3^{3/2}}\left(\frac{\hbar}{m_ec}\right)^2
\sum_{n=1}^\infty\frac{g_n/n}{u(u+1)},
\end{equation}
where $u=n^2E_{e^+}^{\rm kin}/I_{\rm H}$
and $g_n$ is the Gaunt factor for forming the Ps in the energy
state with principal quantum number $n$. The Gaunt factor is close to
unity and can be approximated as \cite{sea}
\begin{equation}
g_n\approx 1+0.1728\left[\frac{u-1}{n^{2/3}(u+1)^{2/3}}\right]
-0.0496\left[\frac{u^2+(4/3)u+1}{n^{4/3}(u+1)^{4/3}}\right].
\end{equation}
The calculation of the cross sections for Ps formation with bound
electrons is rather complex. Here we use the measured cross sections 
$\sigma_{\rm Ps,H}$ and $\sigma_{\rm Ps,He}$ for Ps formation with 
the electrons in the H \cite{zhou} and He \cite{overton} atoms, respectively.
The cross section for direct annihilation with free electrons multiplied
by the number of free electrons per nucleon in the stellar surface
region under consideration, $Y_{e^-}\sigma_{\rm da,f}^{\rm slow}$, 
is compared in Fig.~\ref{fig-xsps} with the corresponding quantities 
$Y_{e^-}\sigma_{\rm Ps,f}$, $Y_{\rm H}\sigma_{\rm Ps,H}$, and
$Y_{\rm He}\sigma_{\rm Ps,He}$ for Ps formation with free electrons
and the electrons in the H and He atoms, respectively.

\end{document}